\documentclass[twocolumn]{aastex63}
\usepackage{lineno}
\usepackage{graphicx}
\usepackage{float} 
\usepackage{subfigure} 
\usepackage{enumerate} 
\usepackage{charter}

\usepackage{tabularx}
\usepackage[flushleft]{threeparttable}
\def\arcsec#1#2{#1$^{\prime\prime}_{^\textrm{.}}$#2}
\shorttitle{The velocity discrepancy between CO and H$\alpha$ rotations}
\shortauthors{Yung-Chau et al.}

\graphicspath{{./}{figures/}}

\begin{document}

\title{The ALMaQUEST survey. VIII. What causes the velocity discrepancy between CO and H$\alpha$ rotation curves in galaxies?}
\author[0000-0001-6106-5383]{Yung-Chau Su}
\affiliation{Department of Physics , National Taiwan University, No. 1, Section 4, Roosevelt Road,  Taipei 10617, Taiwan}
\affiliation{Institute of Astronomy and Astrophysics, Academia Sinica, No. 1, Section 4, Roosevelt Road, Taipei 10617, Taiwan}

\author[0000-0001-7218-7407]{Lihwai Lin}
\affiliation{Institute of Astronomy and Astrophysics, Academia Sinica, No. 1, Section 4, Roosevelt Road, Taipei 10617, Taiwan}

\author[0000-0002-1370-6964]{Hsi-An Pan}
\affiliation{Department of Physics, Tamkang University, No.151, Yingzhuan Rd., Tamsui Dist., New Taipei City 251301, Taiwan}

\author[0000-0003-1045-0702]{Carlos L\'{o}pez Cob\'{a}}
\affiliation{Institute of Astronomy and Astrophysics, Academia Sinica, No. 1, Section 4, Roosevelt Road, Taipei 10617, Taiwan}

\author[0000-0001-5615-4904]{Bau-Ching Hsieh}
\affiliation{Institute of Astronomy and Astrophysics, Academia Sinica, No. 1, Section 4, Roosevelt Road, Taipei 10617, Taiwan}

\author[0000-0001-6444-9307]{Sebasti\'{a}n F. S\'{a}nchez}
\affiliation{Instituto de Astronom\'{i}a, Universidad Nacional Autónoma de M\'{e}xico, Circuito Exterior, Ciudad Universitaria, Ciudad de M\'{e}xico 04510, Mexico}

\author[0000-0003-0080-8547]{Mallory D. Thorp}
\affiliation{Department of Physics \& Astronomy, University of Victoria, Finnerty Road, Victoria, British Columbia, V8P 1A1, Canada}

\author[0000-0003-4980-1012]{Martin Bureau}
\affiliation{Sub-Department of Astrophysics, University of Oxford, Denys Wilkinson Building, Keble Road, Oxford OX1 3RH
}

\author[0000-0002-1768-1899]{Sara L. Ellison}
\affiliation{Department of Physics \& Astronomy, University of Victoria, Finnerty Road, Victoria, British Columbia, V8P 1A1, Canada}

\begin{abstract}
We compare the CO(1-0) and H$\alpha$ kinematics in 34 nearby galaxies, selected from the ALMaQUEST and EDGE-CALIFA surveys. We use 3-D Barolo, a 3D tilted ring model, to derive the CO and H$\alpha$ rotation curves. Before comparing rotation curves in the 34 nearby galaxies, we found systematics between the MaNGA and the CALIFA data using eight MaNGA-CALIFA overlapping galaxies. We assume the rotation curves based on the MaNGA data are accurate and made the corresponding correction to the CALIFA data. Our result shows that $\sim$56\% (19/34) of our galaxies present slower H$\alpha$ rotation curves compared to the CO rotation curves, with a median value of 6.5 km/s. The remaining galaxies (15/34) show consistent CO-H$\alpha$ rotation velocity within uncertainties. As a result, the Ha rotation may underestimate the total dynamical mass by ~6\% for a circular velocity of 200 km/s (the median value in our sample). Furthermore, the velocity difference between the CO and H$\alpha$ rotation velocity is found to correlate with the velocity dispersion difference between CO and H$\alpha$, suggesting that the gas pressure plays a role in the velocity discrepancy. After incorporating the effect of pressure support due to the turbulent gas motion to our sample, the median value of the velocity differences decreases to 1.9 km/s, which in turn reduces the underestimation of dynamical mass to $\sim$2\%. Finally, we also investigate the role that the extra-planar diffuse ionized gas (eDIG) plays in the CO-H$\alpha$ velocity discrepancy.

\end{abstract}



\section{Introduction} \label{sec:intro}
The galaxy rotation curve is the relation between the mean orbital speed of stars or gas in a given galaxy and the radial distance from the galactic center. Practically, the rotation curves in galaxies can be measured by the redshifted emission or absorption lines such as HI 21 cm line (tracing the atomic gas), CO emission line (tracing the molecular gas), H$\alpha$ line (tracing the ionized gas), and the stellar absorption lines (tracing the stars), as described in detail in \cite{2001ARA&A..39..137S}. One of the important applications of rotation curves is the constraint on the dynamical mass distribution, which provides a strong evidence of the presence of dark matter \citep{2001ARA&A..39..137S,2008AJ....136.2648D}. To derive the dynamical mass from rotation curves, one important assumption is that the rotating disk in the galaxy's radial equilibrium is balanced by the centrifugal force and the gravitational force. In other words, it assumes that the rotation velocity we measure is an optimal tracer of the circular velocity $V_c=\sqrt{R\frac{\partial \Phi}{\partial R}}$. This assumption requires the emitting material to be dynamically cold so that we can ignore the pressure term in the Virial theorem \citep{2010ApJ...721..547D,2017MNRAS.466.4159I}.\\ 

Several studies have shown that CO is a good tracer of the circular velocity \citep{2013MNRAS.429..534D,2018MNRAS.477..254L,2020ApJ...897..122L}. However, it seems not the case for the H$\alpha$ emission line. \cite{2013MNRAS.429..534D} compared the CO and ionized rotation curves in 24 early-type galaxies (ETG) from the $ATLAS^{3D}$ survey. They find that 80\% of their sample show slower ionized gas rotation velocities. The other 20\% show consistent CO and ionized rotation within uncertainties. They attribute the systematic velocity discrepancy to the presence of pressure support to the ionized gas. Similar studies have also been carried out for late-type galaxies. \cite{2018ApJ...860...92L} compare the CO and ionized rotation curves in 17 rotational-dominated late-type galaxies, selected from the EDGE-CALIFA survey \citep{2012A&A...538A...8S,2017ApJ...846..159B}. They also find that 75\% of their sample show slower H$\alpha$ rotation velocity than the molecular gas, while the other 25\% show consistent CO and H$\alpha$ rotations within errors. They attribute this to the presence of extra-planar diffuse ionized gas (eDIG). Since the rotation velocity of eDIG is slower than the ionized gas on the mid-plane \citep{2019ApJ...882...84L,2017ApJ...839...87B}, the eDIG could reduce the observed mean rotation velocity, producing the systematic rotation velocity offset between CO and ionized gas. These studies suggest that the CO-H$\alpha$ velocity discrepancy is common in nearby galaxies regardless of morphology types.\\ 

Understanding the reason behind such CO-H$\alpha$ velocity discrepancy is important because the generation of mass models was conventionally done based on H$\alpha$ rotation curves \citep{1997ASPC..117...74B,1999A&AS..137..419O,2001AJ....122.2396D,2008ApJ...676..920K,2009A&A...504..789E,2019ApJ...886..153D} given that it is relatively easier to obtain a larger sample with H$\alpha$ rotation measurement compared to other tracers. The lower velocity in H$\alpha$ rotation implies that these mass models are potentially biased to a lower value. Hence, the comprehensive understanding of CO-H$\alpha$ velocity discrepancy is important to quantify the potential mass bias based on Ha rotations.\\

The ALMA-MaNGA QUEnching and STar Formation (ALMaQUEST;  \citealt{2020ApJ...903..145L}) and EDGE-CALIFA  \citep{2012A&A...538A...8S,2017ApJ...846..159B} surveys, which combined the CO(1-0) observations and integral field unit (IFU) spectroscopy data for 46 and 126 nearby galaxies respectively, provide us a good opportunity to compare CO and H$\alpha$ rotation curves in a considerable amount of sample and also in a wide radial extension. Besides the high spatial and spectral resolution of CO(1-0) and H$\alpha$ emission line provided by these datasets, the spatially matched resolution in CO(1-0) and H$\alpha$ observations make it more ideal to compare the molecular and ionized gas rotation curves. In this study, with the selected 34 nearby galaxies from the combined ALMaQUEST and EDGE-CALIFA dataset, we aim to further investigate the CO-H$\alpha$ velocity discrepancy with a larger sample size.\\

In Section \ref{sec:data}, we describe the data used in this study. Section \ref{sec:method} and \ref{sec:analysis} present the method and tools we use to measure the CO and H$\alpha$ rotation curves. The main results of this study and the interpretations of the CO-H$\alpha$ velocity discrepancy are given in Section \ref{sec:result} and Section \ref{sec:discussion}, respectively. Finally, we present the summary of this study in Section \ref{sec:conclusion}.\\

Throughout this paper, we adopt the following cosmology:$H_0=70\, km\,s^{-1}\,Mpc^{-1}$, $\Omega_m=0.3$ and $\Omega_\Lambda=0.7$.

\section{The Data} \label{sec:data}
    \subsection{The MaNGA survey}
    Mapping Nearby Galaxies at APO (MaNGA; \citealt{2015ApJ...798....7B}) is an integral field unit (IFU) survey as part of the SDSS-IV. With the technique of integral field spectroscopy, MaNGA provides the two-dimensional spatially resolved spectra of 10000 galaxies in the range 0.01\textless z\textless0.15. The spectral coverage of MaNGA is from 3600\AA{} to 10300\AA{} with spectral resolution R$\sim$2000, which corresponds to $\sigma_{inst} \approx$ 70km/s at the wavelength of H$\alpha$. The spatial resolution of MaNGA is \arcsec{2}{5}, corresponding to the physical scale 1.5kpc at the average redshift (range from 1.1 to 5.9 kpc for the MaNGA sample included in this study). The spaxel size in the data products is 0.5$''$. Details about the MaNGA survey can be found in \cite{2015ApJ...798....7B}. 
    The H$\alpha$ data cubes for MaNGA galaxies are based on the SDSS Data Release 15 (DR15) analyzed through the MaNGA reduction pipeline \citep{2016AJ....152...83L}. The H$\alpha$ datacubes used to measure the H$\alpha$ rotation curves for MaNGA galaxies are taken from the MaNGA Data Analysis Pipeline \citep{2019AJ....158..231W}. The emission lines are extracted after subtracting the best-fit stellar continuum from the observed spectra. We only include the spaxels with S/N$>$3 for H$\alpha$ flux in this study. The global SFR, stellar mass and the $\Sigma_*$ of the MaNGA galaxies used in this work are taken from the PIPE3D \citep{2016RMxAA..52...21S,2016RMxAA..52..171S} value-added catalog \citep{2018RMxAA..54..217S}.

     \subsection{The CALIFA survey}
    The Calar Alto Legacy Integral Field Area (CALIFA) survey contains spatially resolved spectroscopy for 667 nearby galaxies (0.005\textless z\textless0.03) \citep{2012A&A...538A...8S,2016A&A...594A..36S}. CALIFA used PPAK IFU \citep{2006PASP..118..129K} on the Calar Alto 3.5m telescope. The whole spectral coverage is from 3400\AA{} to 7300\AA{}, which is covered by two spectral gratings. The lower spectral resolution grating (V500) covers the range from 3745\AA{} to 7300\AA{} with  spectral resolution R$\sim$850, corresponding to $\sigma_{inst} \approx$116km/s at H$\alpha$. The high-resolution grating (V1200) covers the range from 3400\AA{} to 4750\AA{} with spectral resolution R$\sim$1650, i.e., $\sigma_{inst} \approx$75km/s. The typical spatial resolution of CALIFA is \arcsec{2}{5}, which corresponds to $\sim$0.8 kpc at the averaged redshift. More details about the selection procedure and statistical properties of galaxies in the CALIFA survey can be found in \cite{2014A&A...569A...1W}. The H$\alpha$ data cubes for CALIFA galaxies are taken from the PIPE3D Dataproducts \citep{2016RMxAA..52...21S,2016RMxAA..52..171S}. Similar to the processing of the MaNGA H$\alpha$ data, The H$\alpha$ emission lines are extracted after subtracting the best-fit stellar continuum from the observed spectra \citep{2016RMxAA..52...21S,2016RMxAA..52..171S}, and only the spaxels with S/N$>$3 for H$\alpha$ flux are included in this study.
    
    \subsection{The EDGE survey}
    Extragalactic Database for Galaxy Evolution (EDGE) survey \citep{2017ApJ...846..159B} comprises $^{12}$CO(J=1-0; 115.2712 GHz) measurement for 126 nearby galaxies with CARMA in the D and E configurations. The parent sample of EDGE-CALIFA galaxies is observed in E configurations, which consists of 177 galaxies selected based on their infrared brightness and biased to the higher star formation rate (SFR) in the CALIFA sample. The 126 galaxies are then selected and resampled in the D configuration and constitute the final EDGE-CALIFA data. More details about the EDGE-CALIFA survey can be found in \cite{2017ApJ...846..159B}. The spatial resolution of EDGE data is $\sim$\arcsec{4}{5}, corresponding to the physical scale between 0.6 and 2.4 kpc for the EDGE sample included in this study. Since CALIFA and EDGE have different spatial resolutions, with EDGE having the worse resolution, we convolved the CALIFA data to match the EDGE resolution. We note that, in this case, the typical spatial resolution is similar to the spatial resolution of MaNGA data. The spectral resolution is $\sigma_{inst} \approx$20 km/s in the data cube. We use the "dilated" mask to exclude noise (Section 3.1.1 in \citealt{2017ApJ...846..159B}). The mask is created by starting at 3.5$\sigma$ (or greater) peaks in at least two consecutive channel maps in the data cube and then expands down to the surrounding 2$\sigma$ contours. Finally, the additional 1 pixel is added to capture the low-level emission. To investigate the CO-H$\alpha$ velocity discrepancy with a larger sample size, we include the sample used in \cite{2018ApJ...860...92L}, which contains 17 star-forming, late-type galaxies. This set of galaxies are referred to as the EDGE-CALIFA Kinematics Sub-Sample (EDGE-CALIFA KSS hereafter).

    \subsection{The ALMaQUEST survey}
    The ALMaQUEST Survey contains $^{12}$CO(J=1-0; 115.2712 GHz) MaNGA follow-up observation for 46 galaxies with ALMA in the C43-2 configuration. The field of view for each galaxy is $\sim$\arcsec{50}{} with spatial resolution $\sim$\arcsec{2}{5}, which matches the MaNGA resolution. The spectral resolution of $^{12}CO(1-0)$ for ALMaQUEST galaxies is $\sigma_{inst}\approx$11 km/s. To investigate how gas drives the location and evolution of galaxies on the star-forming main sequence, the ALMaQUEST survey covers a broad range of specific star formation rate, spanning from the green valley, main sequence, to the starburst regimes. More details in the ALMaQUEST survey can be found in \cite{2020ApJ...903..145L}. In this study, we only include spaxels with S/N$>$3 for $^{12}$CO(1-0).
    
\begin{deluxetable*}{cccccccccccc}
    \tablecaption{\label{tbl:ALMa_KSS} Parameters for ALMaQUEST KSS}
    \tablehead{  \colhead{Name} & \colhead{z\tablenotemark{a}} & \colhead{Sersic index\tablenotemark{a}} &  \colhead{$log_{10}(M_*)$\tablenotemark{b}} &  \colhead{$log_{10}(SFR)$\tablenotemark{b}} & \colhead{$log(sSFR)$\tablenotemark{b}} & \colhead{Inc\tablenotemark{c}} & \colhead{PA\tablenotemark{c}}
 &\colhead{CO $V_{max}$} &\colhead{$\Delta V$} &\colhead{$\sigma_{CO}$\tablenotemark{d}} & \colhead{$\sigma_{H\alpha}$\tablenotemark{e}}\\ & & & \colhead{($M_\odot$)} & \colhead{(${M_\odot yr^{-1}}$)} & \colhead{({$yr^{-1}$})} &\colhead{($^\circ$)} & \colhead{($^\circ$)} & \colhead{(km/s)} & \colhead{(km/s)} & \colhead{(km/s)} & \colhead{(km/s)} }
    \startdata
7815-12705 & 0.0293 & 1.1 & 10.77 & 0.8 & -10.0 & 67 & 319 & 189.7$\pm$2.0 & 0.6$\pm$1.7 & 14.9$\pm$1.0 & 35.7$\pm$5.4 \\
7977-3704 & 0.027 & 3.2 & 10.45 & -0.59 & -11.0 & 48 & 221 & 193.0$\pm$1.5 & 11.2$\pm$3.0 & 1.2$\pm$1.5 & 31.8$\pm$8.8\\
7977-12705 & 0.0269 & 4.8 & 11.02 & 0.66 & -10.4 & 50 & 210 & 276.7$\pm$5.5 & 24.0$\pm$19.6 & 16.2$\pm$2.5 & 64.2$\pm$6.8\\
8077-9101 & 0.0429 & 2.2 & 10.42 & -0.08 & -10.5 & 76 & 51 & 190.9$\pm$0.9 & 5.2$\pm$6.0 & 9.8$\pm$1.3 & 39.0$\pm$6.8\\
8078-6103 & 0.0282 & 1.9 & 10.8 & 0.51 & -10.3 & 51 & 165 & 215.6$\pm$2.8 & -3.6$\pm$5.7 & 15.7$\pm$1.3 & 32.2$\pm$6.7\\
8078-12701 & 0.0267 & 3.5 & 11.12 & 0.37 & -10.8 & 65 & 195 & 256.5$\pm$3.8 & 8.4$\pm$17.8 & 18.1$\pm$0.8 & 32.7$\pm$4.4\\
8081-9101 & 0.0282 & 1.8 & 10.73 & 0.41 & -10.3 & 71 & 289 & 214.5$\pm$4.1 & 4.9$\pm$6.1 & 17.3$\pm$1.1 & 45.2$\pm$6.2\\
8081-9102 & 0.0336 & 2.0 & 10.81 & 0.17 & -10.6 & 74 & 100 & 239.8$\pm$3.7 & 15.6$\pm$7.6 & 17.9$\pm$1.1 & 44.5$\pm$5.7\\
8082-12701 & 0.0268 & 2.9 & 10.53 & 0.07 & -10.5 & 51 & 9 & 179.0$\pm$1.4 & 4.9$\pm$2.9 & 2.4$\pm$0.8 & 36.8$\pm$4.4\\
8082-12704 & 0.1317 & 5.3 & 11.66 & -0.31 & -12.0 & 62 & 163 & 322.2$\pm$5.0 & 9.1$\pm$3.9 & 15.2$\pm$1.9 & 45.9$\pm$8.8\\
8084-12705 & 0.0252 & 1.9 & 10.6 & -0.06 & -10.7 & 74 & 62 & 181.9$\pm$2.3 & 13.3$\pm$5.0 & 11.4$\pm$1.2 & 35.1$\pm$6.2\\
8086-9101 & 0.0398 & 6.0 & 11.12 & -0.09 & -11.2 & 63 & 265 & 256.1$\pm$1.0 & 3.2$\pm$2.9 & 3.4$\pm$1.1 & 39.0$\pm$5.7\\
8615-9101 & 0.0332 & 2.0 & 10.78 & 0.01 & -10.8 & 74 & 351 & 218.5$\pm$3.2 & 11.8$\pm$5.6 & 13.9$\pm$1.2 & 36.6$\pm$6.1\\
8615-12702 & 0.0207 & 1.2 & 10.21 & 0.1 & -10.1 & 75 & 318 & 102.0$\pm$1.4 & 1.6$\pm$3.3 & 8.2$\pm$1.1 & 28.1$\pm$5.7\\
8618-9102 & 0.0431 & 0.7 & 10.47 & 0.42 & -10.0 & 73 & 327 & 159.5$\pm$3.1 & 5.7$\pm$1.2 & 6.3$\pm$1.9 & 31.8$\pm$7.6\\
8655-12705 & 0.0452 & 2.4 & 10.51 & -1.55 & -12.1 & 58 & 199 & 205.3$\pm$2.7 & 58.7$\pm$2.1 & 5.3$\pm$1.8 & 48.0$\pm$8.9\\
8952-12701 & 0.0282 & 1.6 & 10.78 & -0.66 & -11.4 & 58 & 27 & 218.9$\pm$2.2 & 6.3$\pm$8.3 & 0.7$\pm$1.5 & 29.3$\pm$5.8\\
    \enddata
    \tablenotetext{a} {From SDSS nsa catalog.}
    \tablenotetext{b} {From \cite{2020ApJ...903..145L}}
    \tablenotetext{c} {3DB fit result.}
    \tablenotetext{d} {The radial averaged velocity dispersion derived from 3DB fit result. The region within one beam from the center is excluded from the calculation. For 8952-12701, which is lack of CO detection in the central region, We start the sampling from four beams away from the center.}
    \tablenotetext{e} {The radial averaged velocity dispersion derived from 3DB fit result. The region within one beam from the center is excluded from the calculation.}
\end{deluxetable*}


\begin{deluxetable*}{cccccccccccc}
    \tablecaption{\label{tbl:EDGE_CALIFA_KSS} Parameters for EDGE-CALIFA KSS}
    \tablehead{  \colhead{Name} & \colhead{z\tablenotemark{a}} & \colhead{Sersic index\tablenotemark{a}} &  \colhead{$log_{10}(M_*)$\tablenotemark{b}} &  \colhead{$log_{10}(SFR)$\tablenotemark{b}} & \colhead{$log(sSFR)$\tablenotemark{b}} & \colhead{Inc\tablenotemark{c}} & \colhead{PA\tablenotemark{c}}
 &\colhead{CO $V_{max}$} &\colhead{$\Delta V$} &\colhead{$\sigma_{CO}$\tablenotemark{d}} & \colhead{$\sigma_{H\alpha}$\tablenotemark{e}}\\ & & & \colhead{($M_\odot$)} & \colhead{(${M_\odot yr^{-1}}$)} & \colhead{({$yr^{-1}$})} &\colhead{($^\circ$)} & \colhead{($^\circ$)} & \colhead{(km/s)} & \colhead{(km/s)} & \colhead{(km/s)} & \colhead{(km/s)} }
    \startdata
IC1199 & 0.0164 & 0.8 & 10.78 & 0.1 & -10.7 & 69 & 160 & 198.6$\pm$3.5 & 7.1$\pm$20.8 & 5.4$\pm$2.1 & 32.5$\pm$7.9 \\
NGC2253 & 0.012 & 1.6 & 10.81 & 0.11 & -10.7 & 41 & 120 & 190.0$\pm$9.2 & -1.4$\pm$11.2 & 3.6$\pm$2.0 & 28.3$\pm$8.1 \\
NGC2347 & 0.0146 & 2.3 & 11.04 & 0.1 & -10.9 & 50 & 9 & 287.3$\pm$4.5 & 15.5$\pm$4.5 & 5.5$\pm$2.2 & 44.2$\pm$8.7 \\
NGC2410 & 0.0157 & 2.9 & 11.03 & 0.1 & -10.9 & 72 & 37 & 253.2$\pm$5.8 & 21.2$\pm$9.1 & 5.9$\pm$1.9 & 49.9$\pm$6.6 \\
NGC3815 & 0.0126 & 1.6 & 10.53 & 0.09 & -10.4 & 64 & 245 & 191.6$\pm$5.8 & 10.1$\pm$3.4 & 6.6$\pm$2.2 & 34.6$\pm$8.8 \\
NGC4047 & 0.0114 & 1.5 & 10.87 & 0.1 & -10.8 & 41 & 286 & 219.2$\pm$2.8 & 6.7$\pm$2.3 & 5.0$\pm$2.2 & 26.7$\pm$8.7 \\
NGC4644 & 0.0165 & 1.5 & 10.68 & 0.11 & -10.6 & 66 & 237 & 188.4$\pm$6.3 & 12.9$\pm$5.1 & 6.2$\pm$2.2 & 38.6$\pm$8.8 \\
NGC4711 & 0.0136 & 0.9 & 10.58 & 0.09 & -10.5 & 64 & 40 & 161.9$\pm$2.3 & -3.7$\pm$1.8 & 2.8$\pm$1.8 & 26.0$\pm$7.0 \\
NGC5016 & 0.0087 & 1.3 & 10.47 & 0.09 & -10.4 & 43 & 239 & 177.2$\pm$6.0 & -3.3$\pm$3.7 & 5.2$\pm$2.0 & 22.7$\pm$8.0 \\
NGC5480 & 0.0065 & 1.0 & 10.18 & 0.08 & -10.1 & 45 & 3 & 124.1$\pm$5.8 & -1.4$\pm$5.2 & 3.8$\pm$1.5 & 30.3$\pm$6.2 \\
NGC5520 & 0.0062 & 3.2 & 10.07 & 0.11 & -10.0 & 53 & 65 & 170.4$\pm$3.7 & 19.3$\pm$6.2 & 5.2$\pm$2.1 & 35.1$\pm$8.0 \\
NGC5633 & 0.0079 & 0.8 & 10.4 & 0.11 & -10.3 & 48 & 197 & 176.8$\pm$5.0 & -4.5$\pm$3.1 & 6.7$\pm$1.8 & 25.5$\pm$7.5 \\
NGC5980 & 0.0137 & 1.3 & 10.81 & 0.1 & -10.7 & 69 & 195 & 221.9$\pm$4.5 & 2.7$\pm$17.6 & 8.1$\pm$2.1 & 35.3$\pm$6.2 \\
UGC04132 & 0.0174 & 0.8 & 10.94 & 0.12 & -10.8 & 68 & 29 & 257.6$\pm$5.7 & 10.6$\pm$22.0 & 14.1$\pm$1.9 & 47.0$\pm$6.5 \\
UGC05111 & 0.0224 & 2.0 & 10.82 & 0.12 & -10.7 & 73 & 300 & 232.3$\pm$4.1 & 8.8$\pm$9.6 & 10.7$\pm$1.8 & 41.0$\pm$6.5 \\
UGC09067 & 0.0263 & 1.5 & 10.96 & 0.12 & -10.8 & 64 & 194 & 216.0$\pm$7.6 & -1.4$\pm$6.7 & 5.6$\pm$2.2 & 28.1$\pm$8.6 \\
UGC10384 & 0.0167 & 1.3 & 10.33 & 0.14 & -10.2 & 73 & 91 & 185.1$\pm$5.8 & 6$\pm$4.3 & 16.4$\pm$1.9 & 37.8$\pm$8.0 \
\enddata
    \tablenotetext{a} {From SDSS nsa catalog.}
    \tablenotetext{b} {From \cite{2017ApJ...846..159B}}
    \tablenotetext{c} {3DB fit result.}
    \tablenotetext{d} {The radial averaged velocity dispersion derived from 3DB fit result. The region within one beam from the center is excluded from the calculation.}
    \tablenotetext{e} {The radial averaged velocity dispersion derived from 3DB fit result. The region within one beam from the center is excluded from the calculation.}
\end{deluxetable*}

\section{Method} \label{sec:method}
In this section, we introduce the general methodology for both the ALMaQUEST KSS and the EDGE-CALIFA KSS.
\subsection{Tilted-ring model}
Tilted ring model \citep{1974ApJ...193..309R, 1985ApJ...295..305V, 1989A&A...223...47B} is a technique widely used to describe the kinematics of spiral galaxies. The main assumptions of this technique are the following: (i) The emitting material is settled in a thin disk, (ii) the rotation velocity of the material only depends on the distance from the center, and (iii) the motion of the material is dominated by circular motion. Under these assumptions, the tilted ring model breaks the rotating disk into a series of independent concentric rings with different radii, each of which has its geometric and kinematics parameters. Since the rotating disk is not necessarily face-on from the line of sight, the projection of each ring on the sky is usually parameterized as an elliptical ring. For a given projected ring, the projected velocity along the line of sight $V_{los}$ is given by:

\begin{equation} \label{eq:1}
V_{los}=V_{sys}+V_{rot}(R)cos(\theta)sin(i),
\end{equation}

where $V_{sys}$ is the system velocity of the galaxies, $V_{rot}(R)$ is the deprojected rotation velocity of the ring with deprojected radius R, $\theta$ is the deprojected position angle ($0^{\circ}$ is at the major axis of the projected elliptical rings), and i is the inclination between the line of sight and the rotating disk ($0^{\circ}$ is face-on). The standard approach to apply the tilted ring model was to fit the 2D-velocity maps, but some drawbacks of such a 2D fitting algorithm exist. The most severe one is the beam smearing effect, which tends to flatten the gradient of the velocity field \citep{2000AJ....119.1579V,2009A&A...493..871S}. In this work, we adopt an alternative approach by using a 3D-fitting software based on the concept of the tilted ring model called $^{3D}BARALO$ (3DB hereafter; \citealt{2015MNRAS.451.3021D}) to measure the kinematics and geometric parameters.

\subsection{3D-Barolo}
3DB \citep{2015MNRAS.451.3021D} adopts a 3D tilted ring model that fits the 3-D datacube (two axes define sky coordinate and one axis defines spectrum direction) directly. In practice, 3DB fits the datacube ring by ring by building an artificial 3D model cube and finds the best fit through minimizing the residual calculated pixel by pixel between the datacube and the model cube. Before comparing the datacube and the model cube, the former need to be convolved to the same spatial and spectral resolution as the latter. This ensures the full control of instrumental effects, such as the beam smearing effect, which can significantly influence the fitting result at the region where the velocity gradient is large.\\

The residual function to be minimize in this study can be expressed as

\begin{equation}\label{eq:r1}
F=\frac{1}{n}\sum_{i=1}^{n} \Delta r_i w(\theta_i),
\end{equation}

where n is the number of pixels, $\Delta r$ is the residual at each pixel, $w(\theta)$ is the weighting function of azimuthal angle. $\Delta r$ and the weighting function can be expressed as

\begin{equation}\label{eq:r1_2}
\Delta r=\mid M-D \mid
\end{equation}
and
\begin{equation}\label{eq:r1_3}
w(\theta)=\mid\cos(\theta)\mid^2,
\end{equation}

where M and D are the flux values of the model and the data, respectively. $\theta$ is the position angle ($\theta=0^{\circ}$ for the major axis). We note that there is no intensity weighting in the residual function. Hence, the faint pixels carry the same weights as bright ones.\\

In 3DB, each ring with deprojected radius R and width W is characterized by 8 parameters: Spatial coordinate of the centre ($x_0$, $y_0$),  $V_{sys}$, $V_{rot}$, velocity dispersion ($\sigma_{v}$), i, PA (the position angle between the North and the major axis, represented in degree from the North in the counterclockwise direction), gas component surface density ($\Sigma$) and ring thickness along the axisymmetric axis ($z_d$).\\

In this work, we focus on the kinematics parameters $V_{rot}$ and $\sigma_v$. To reduce the parameter space, some assumptions are made on other parameters before performing the fitting. First, we exclude the gas surface density from the fit by normalizing the flux in the model cube to that in the original data cube. Two different kinds of normalization are provided in 3DB: pixel-by-pixel and azimuthally averaged. In the former case, the integrated flux of each pixel in the model cube is normalized to that in the data cube. In the second case, the model is normalized to the azimuthal-averaged flux in each ring. More details can be found in \cite{2015MNRAS.451.3021D}. In this study, we apply the pixel-by-pixel approach since it allows us to accounts for the non-axisymmetric gas distribution. On the other hand, we fix the disk scale height to 100 pc. In Sec \ref{sec:sec6.2}, we discuss the impact about this assumption. Finally, we assume a constant inclination, position angle, systematic velcity and kinematic center for each galaxy.

\section{Analysis} \label{sec:analysis}
\subsection{Fit rotation curves}
To derive reliable rotation curves from the tilted ring model, it is crucial to set initial parameters properly. The initial dynamical center is the photometric center and the initial values for inclinations and PAs are derived from the photometric fit from NASA-Sloan Atlas (NSA) catalog \citep{2011AJ....142...31B}. The initial $V_{sys}$ is calculated as the midpoint between the velocities corresponding to the 20$\%$ of the peaks of the global line profile \citep{2015MNRAS.451.3021D}. We assume that all rings in each galaxy share the same dynamical center, inclination, PA, and $V_{sys}$. The initial values of rotation velocity and velocity dispersion are set to be 100 and 30 km/s, respectively. \\

We apply a two-step process to fit the rotation curves. In the first step, we use one unit of spatial resolution as the annulus width and fit the geometric parameters, rotation curves, and velocity dispersion at the same time. The purpose of the first step is to obtain reliable geometric parameters. In the second step, we fix the geometric parameters and fit the rotation velocity and velocity dispersion with half of the unit of the spatial resolution as the annulus width. These two steps ensure an accurate rotation curve measurement.\\

In the new version, we quantify the errors by 3DB’s method for both the rotation velocity and the velocity dispersion. After finding the best model by minimizing the residual function, 3DB normalizes the value of the minimum of the residual function. Then, 3DB calculates a number of models by a Gaussian distribution oversampling. The center of the Gaussian distribution is the point that minimizes the residual function in the parameter space. Errors for each parameter are determined by the range where the residual increases 5 percent with respect to the minimum. i.e., the range where the residual function is smaller than 1.05. More details can be found in \cite{2015MNRAS.451.3021D}.\\

When measuring the rotation curves in galaxies, one of the important issues is the beam smearing effect, which causes the degeneracy between rotation velocity and velocity dispersion, and flattens the velocity gradient (e.g., \citealt{2000AJ....119.1579V,2009A&A...493..871S}). The beam smearing effect becomes significant in the case where the velocity gradient is large, which is usually in the central region for a galaxy. Although 3DB allows better control of such instrumental effect compared to the 2D-fitting methods, some issues were found when measuring the central rotation curves, which will be discussed in more detail in Sec.\ref{sec:sec5.1}. The H$\alpha$ rotation curves for our sample are fit using the same method under the assumption that CO and H$\alpha$ are co-planar. In other words, CO and H$\alpha$ share the same kinematics center, inclination, and PA.\\

\subsection{The subsample selected for this study} \label{sec:4.2}
To reliably measure rotation curves by the tilted ring model, there are practical upper and lower limits to the inclination of galaxies, as discussed in \cite{2008AJ....136.2648D}. For high-inclination galaxies, it is difficult to derive rotation curves since the line-of-sight penetrates several layers with different distances from the center. For low-inclination galaxies, the decreased projected component of rotation velocity makes it difficult to derive reliable rotation curves. As adopted by \cite{2008AJ....136.2648D}, the practical upper and lower limits of inclination are suggested to be $40^\circ$ and $75^\circ$, respectively. We further exclude galaxies that show apparent bar structures, interacting galaxies, and galaxies with distorted morphology through visual inspection, as similarly done by \cite{2020ApJ...900..109A}. This selection makes sure that the gas motion is dominated by rotation and also validates the assumption of constant inclination and PA. After applying these selection criteria to the ALMaQUEST galaxies, 17 galaxies are selected and referred to as the ALMaQUEST Kinematic Sub-Sample (ALMaQUEST KSS hereafter). We note that the 17 galaxies from the EDGE-CALIFA KSS fit well to these selection criteria. The result of this research is based on the study on the combined 17 ALMaQUEST KSS and the 17 galaxies from the EDGE-CALIFA KSS.

\section{Result} \label{sec:result}
\subsection{The systematic discrepancy between CO and \texorpdfstring{H$\alpha$}{} rotation} \label{sec:sec5.1}

\begin{figure*}[htbp]
    \centering 
    \includegraphics[width=0.9\textwidth]{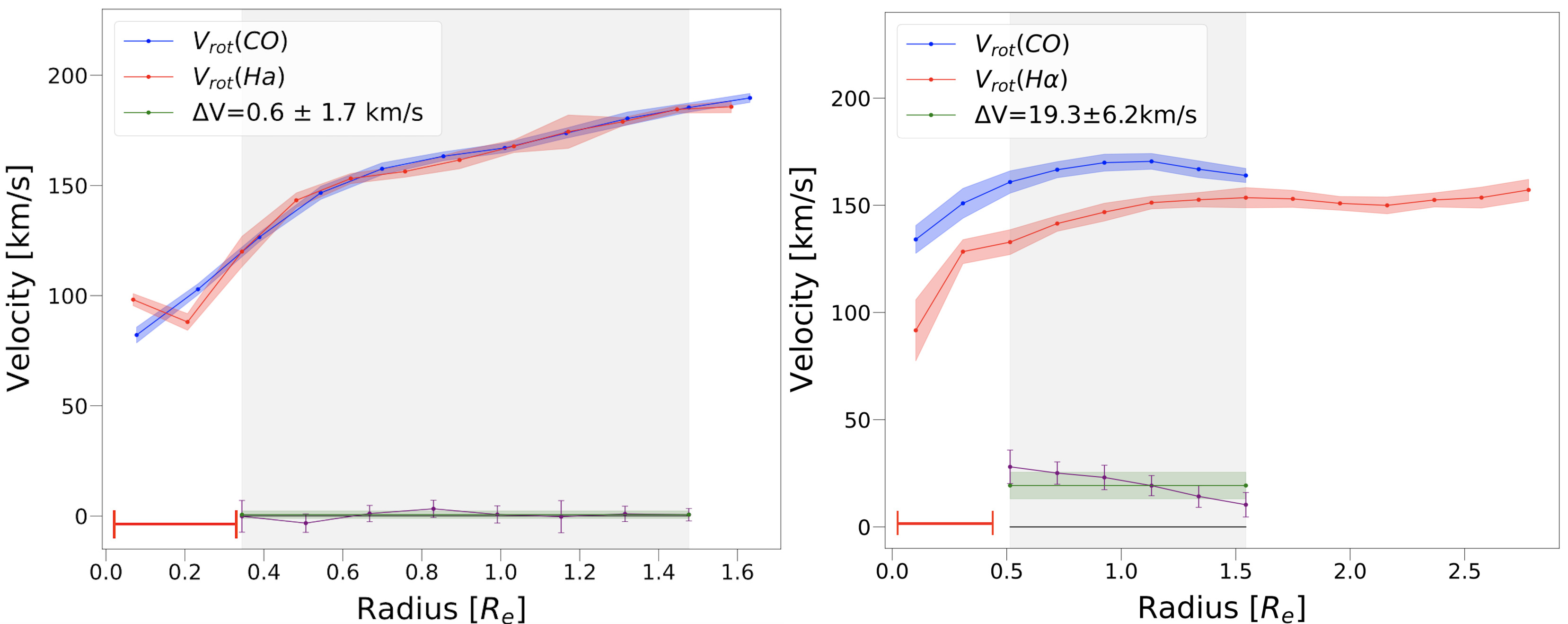}
    \caption{The left panel shows the rotation curves of one of our galaxies from the ALMaQUEST survey, 7815-12705. Blue one stands for $V_{rot}(CO)$ and the red one is $V_{rot}(H \alpha)$. $\Delta V$, the variance weighted mean of the rotation difference between $V_{rot}(CO)$ and $V_{rot}(H \alpha)$, is calculated within the gray area and is presented as the green line. The purple line shows the velocity differences in each radius. The error of $\Delta V$ is the standard deviation of the velocity differences in sample points. The shadow areas present the errors of each quantity. The red bar at the lower-left corner is the beam size. The right panel shows the rotation curves in NGC5520, one of the galaxies in the EDGE-CALIFA survey which shows CO-H$\alpha$ velocity discrepancy. We note that, in general, the radial range where we compare the CO and H$\alpha$ rotation curves are similar between the ALMaQUEST samples and the EDGE-CALIFA samples, which covers from $\sim 0.3 R_e$ to $\sim 1.5 R_e$.}
    \label{fig:rotation_curve}
    \end{figure*}

Previous studies in the literature have found that H$\alpha$ tends to rotate either consistently or slower than CO \citep{2018ApJ...860...92L, 2013MNRAS.429..534D,2005ApJ...621..757S}. In this study, we show that the CO-H$\alpha$ velocity discrepancy is seen in both ALMaQUEST KSS and the EDGE-CALIFA KSS. The left panel in Figure \ref{fig:rotation_curve} shows an example selected from the ALMaQUEST survey with consistent CO and H$\alpha$ rotation. As a comparison, the right panel shows a galaxy from the EDGE-CALIFA survey which has an apparent velocity discrepancy between CO and H$\alpha$ rotation. The same plots for all the galaxies included in this study can be found in Figure \ref{fig:ALMa_mom_rot_vdisp_1} and Figure \ref{fig:EDGE_mom_rot_vdisp_1}. We note that, although the galaxies in the CALIFA survey are covered up to 2.5 effective radius ($R_e$; \citealt{2014A&A...569A...1W}), which is larger than that of MaNGA galaxies (1.5$R_e$; \citealt{2015ApJ...798....7B}), the radial range in which we compare the CO and H$\alpha$ rotation curves are set to be between 0.3$R_e$ and 1.5$R_e$ for both the ALMaQUEST and the EDGE-CALIFA samples that are constrained by the coverage of CO. On the other hand, as we can see, there is a turnover feature shown in H$\alpha$ rotation curves in the central region. Such a feature is commonly found in the H$\alpha$ rotation curves among our samples. To test if this feature is artificial, we degrade the spectral resolution of our CO datacubes, which have original spectral resolution $\sigma_{inst}\approx$11km/s, to the spectral resolution of MaNGA H$\alpha$ data ($\sigma_{inst}\approx$70km/s) by convolving them with a Gaussian kernel along the spectral axis. After fitting the convolved data by 3DB, we found such a feature also appears in the CO rotation curves. This phenomenon might be caused by the following two effects: (i) it is more difficult to derive reliable rotation velocity and velocity dispersion at the same time with a low spectral resolution; (ii) there are fewer data points in the most central region, which make it difficult to construct a reliable fit result. In this study, we exclude the rotation curves fit result within one resolution area from the center in the following analysis. \\

To quantify the velocity difference between CO and H$\alpha$ rotation curves for a given galaxy, we first linearly interpolate the H$\alpha$ rotation curve and resample the H$\alpha$ rotation velocity at the same radii where we sample the CO rotation velocity. Then, for each galaxy, we calculate the velocity difference between CO and H$\alpha$ rotation at each radius. We define $\Delta V$ for an individual galaxy as the variance weighted mean of the velocity differences $V_{rot,CO}-V_{rot,H\alpha}$. $\sigma_{\Delta V}$, the error of $\Delta V$, is defined as the standard deviation of the velocity differences. In this study, we say that a galaxy has CO-H$\alpha$ velocity discrepancy if $\Delta V$ \textgreater $\sigma_{\Delta V}$ and has consistent CO-H$\alpha$ rotation if $\Delta V \leq \sigma_{\Delta V}$. \\

As the spectral resolution differs between MaNGA and CALIFA datacubes, it is important to quantify whether there exists systematics or not in the measurement of the H$\alpha$ rotation curves due to the difference in the instrumental characteristics. To conduct this test, we select in total 8 galaxies that are included in the main samples of both MaNGA and CALIFA surveys \footnote{These 8 galaxies are independently selected and are not part of the ALMaQUEST nor the EDGE-CALIFA samples.}. When comparing the rotation curves of the 8 MaNGA-CALIFA overlapping galaxies, we found systematics between the rotation curves fitted from the MaNGA H$\alpha$ datacubes and that fitted from the CALIFA H$\alpha$ datacubes. The latter tend to have slower rotation velocity. We adopt the definition of $\Delta V$ and use it to quantify the velocity differences between the MaNGA H$\alpha$ and the CALIFA H$\alpha$ rotation curves in these 8 overlapping galaxies (labeled as $\Delta V_{MC}$). The median value of $\Delta V_{MC}$ in these 8 galaxies is 9.8 km/s. Finally, we use bootstrap to test if this systematics is reliable, and a median value of $\Delta V = 8.5$ with a standard deviation of 1.3 km/s is found. In the appendix \ref{appen:MC_test}, we discuss the details of this test. This systematics causes $\Delta V$ in the EDGE-CALIFA KSS to become larger (or smaller in the ALMaQUEST KSS). To compensate for this systematics, we subtracted the $\Delta V$ in the CALIFA KSS by 8.5 km/s. All the discussions afterward have included this correction.\\

\begin{figure}[tbp] 
    \centering 
    \includegraphics[width=0.5\textwidth]{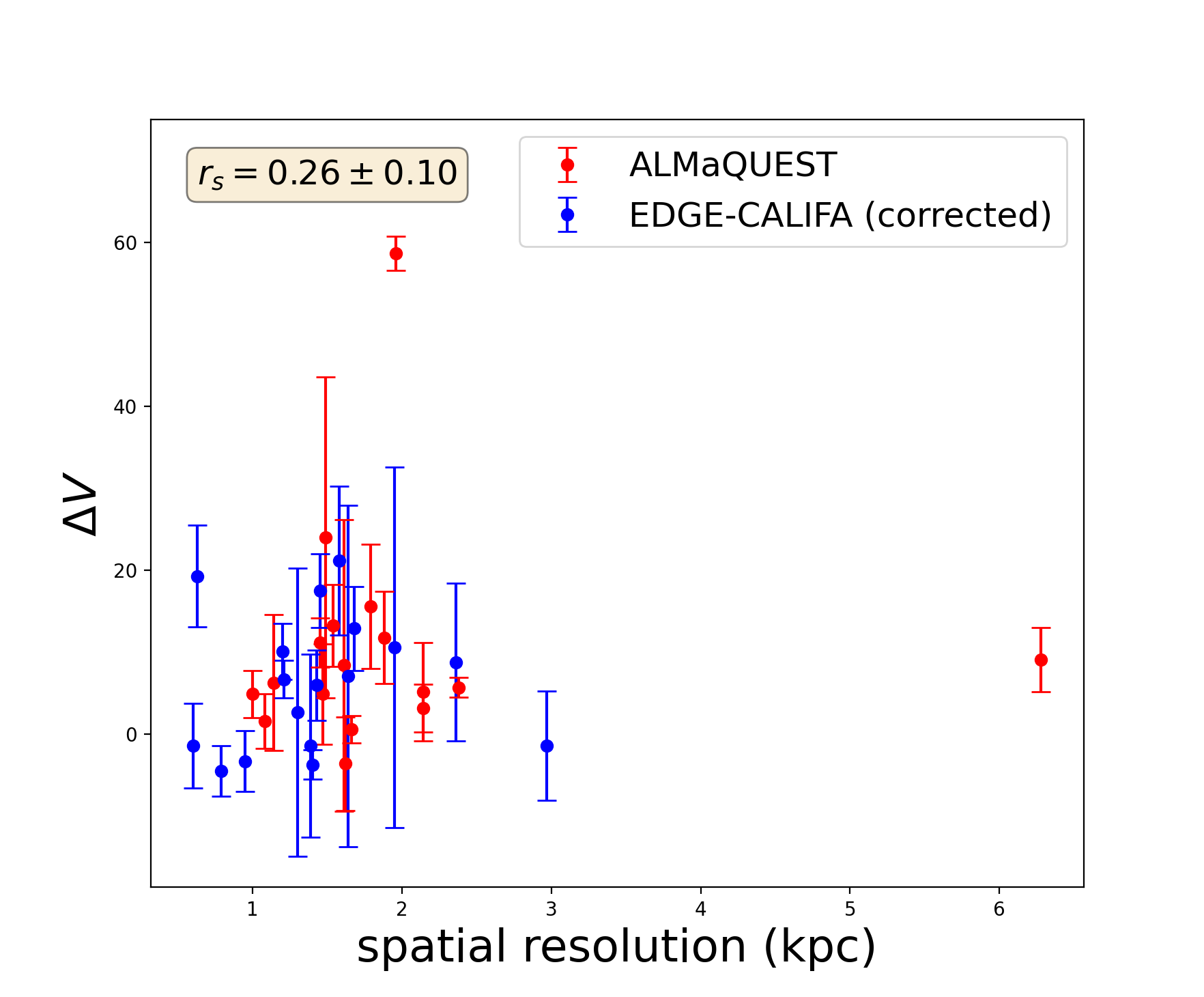}
    \caption{The relation between the $\Delta V$ and the physical spatial resolution for the ALMaQUEST KSS (red points) and the EDGE-CALIFA (blue points).}
    \label{fig:DeltaV2spa_res}
    \end{figure}

On the other hand, given the fact that the physical spatial resolutions of the ALMaQUEST KSS and the EDGE-CALIFA KSS are similar to each other, it is unlikely that the spatial resolution would raise an issue about the systematics between these two datasets. Figure \ref{fig:DeltaV2spa_res} shows the relation between the $\Delta V$ and the physical spatial resolutions. As we can see, the physical spatial resolutions of the two datasets are comparable. Moreover, the lack of trend between the $\Delta V$ and the physical spatial resolutions are also found. The same results are found for the case which does not apply the MaNGA-CALIFA systematics correction to the EDGE-CALIFA KSS.\\

Based on the definition of $\Delta V$, $\sim$56\% (19/34) of the sample shows higher CO rotation velocity, while $\sim$44\% (15/34) of the sample show consistent CO-H$\alpha$ rotation. Some important parameters from 3DB fit result are listed in the table \ref{tbl:ALMa_KSS} and table \ref{tbl:EDGE_CALIFA_KSS}. \\

\begin{figure}[tbp] 
    \centering 
    \includegraphics[width=0.5\textwidth]{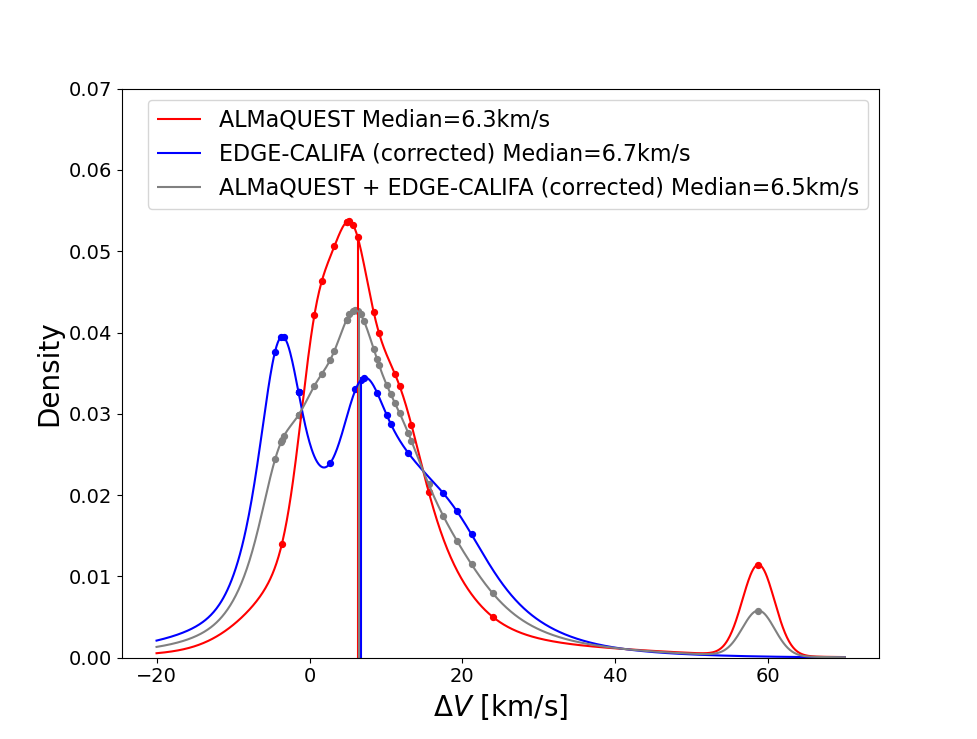}
    \caption{Kernel density distribution of $\Delta V$. Each galaxy is presented as a normalized Gaussian kernel with dispersion $\sigma_{\Delta V}$ and centered at $\Delta V$. The gray curve shows the $\Delta V$ kernel density distribution (KDD) in our whole sample. The red curve shows the KDD for the ALMaQUEST KSS. The blue curve shows the KDD for the EDGE-CALIFA KSS after correcting for the MaNGA-CALIFA systematics}.
    \label{fig:kernel_density}
    \end{figure}

Figure \ref{fig:kernel_density} shows the normalized $\Delta V$ kernel density distribution (KDD) of our galaxies. In this figure, we use a normalized Gaussian kernel to represent a galaxy, which is centered at $\Delta V$ with the dispersion $\sigma_{\Delta V}$. After summing up all the Gaussians, the total distribution histogram is normalized to the unit area again. As we can see, the KDD shows that CO tends to have a higher rotation speed. The median $\Delta V$ of the ALMaQUEST+EDGE-CALIFA KSS galaxies is 6.5 km/s. In general, The distribution of $\Delta V$ in EDGE-CALIFA KSS (after correcting for the MaNGA-CALIFA systematics described in Appendix \ref{appen:MC_test}) is similar in compared to the ALMaQUEST KSS. \\

\begin{figure}[tbp] 
    \centering 
    \includegraphics[width=0.5\textwidth]{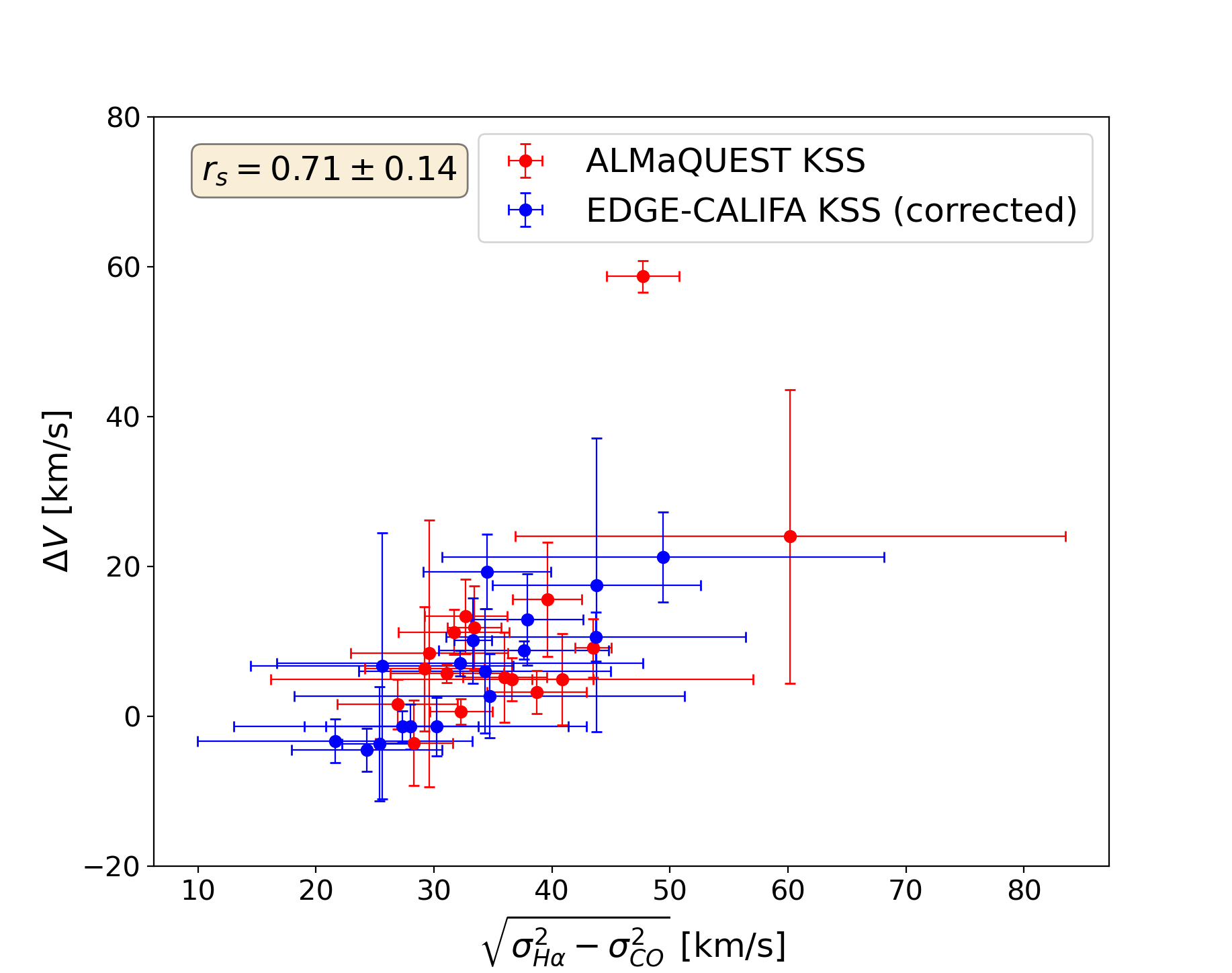}
    \caption{The rotation velocity difference between CO and H$\alpha$ ($\Delta V$) v.s. the radial mean of $\sqrt{\sigma_{H\alpha}^2-\sigma_{CO}^2}$ for the ALMaQUEST KSS (red points) and the EDGE-CALIFA KSS (blue) samples. The error of the velocity dispersion difference is characterized by the standard deviation of the measurements in the annuli. The Spearman correlation coefficient ($r_s$) shown in the upper-left corner is for the combined sample, which yields a strong correlation. The $r_s$ for the ALMaQUEST KSS and the EDGE-CALIFA KSS are $0.51\pm0.19$ and $0.86 \pm 0.22$, respectively.The positive correlation between $\Delta V$ and the velocity dispersion quadrature suggests that the gas pressure contributes to the CO-H$\alpha$ velocity discrepancy.}
    \label{fig:deltaV_Vdisp}    
    \end{figure}

Figure \ref{fig:deltaV_Vdisp} shows the relation between the $\Delta V$ and the radial mean of  $\sqrt{\sigma_{H\alpha}^2-\sigma_{CO}^2}$ for each individual galaxy. The velocity dispersion differences are calculated within the region where $\Delta V$ is calculated. As we can see, there is a positive trend between these two parameters. We run the Spearman correlation analysis and find that the correlation coefficients give $r_s$ = 0.51 $\pm$  0.19, 0.86 $\pm$ 0.22, and 0.71 $\pm$ 0.14 for the ALMaQUEST KSS, EDGE-CALIFA KSS, and the combined sample, respectively. Since the magnitude of the velocity dispersion is related to the magnitude of pressure gradient in the radial direction \citep{2010ApJ...721..547D,2018ApJ...860...92L,2013MNRAS.429..534D}, this gives us a hint that the radial pressure support may be responsible for the CO-Ha velocity discrepancy. We will discuss this scenario in the Sec.\ref{sec:sec5.2}.\\

We also study the dependence of the $\Delta V$ on other global parameters, such as star formation rate (SFR), stellar mass ($M_{*}$), and specific star formation rate (sSFR), morphology, inclination, CO $V_{max}$, as well as on the $\Sigma_{SFR}$ and $\Sigma_{M_{*}}$ within the area where the $\Delta V$ is calculated. These plots are shown in Figure \ref{fig:deltaV_other_parameters}. Except for the Sersic index, we did not find any apparent trend between $\Delta V$ and other parameters. These plots suggest that (i) no correlation be found between the $\Delta V$ and the inclination; (ii) $\Delta V$ tends to be smaller in late-type galaxies (our galaxies are biased to the late-type, though).


\subsection{The radial pressure support} \label{sec:sec5.2}
Under the condition that gas motion is in equilibrium, the relation between the gravitational potential ($\Phi$) and the $V_{rot}$ can be expressed by the following equation\citep{2017MNRAS.466.4159I}:

\begin{equation} \label{eq:2}
\frac{1}{\rho}\frac{\partial P}{\partial R}=
-\frac{\partial \Phi}{\partial R}+\frac{V^2_{rot}}{R}
\end{equation}

Where $\rho$ is the volumetric density of the gas, and P is the gas pressure. Under the assumption that turbulence pressure dominates the pressure term, the pressure term can be expressed as $\rho \sigma^{2}_{v}$ \citep{2010ApJ...721..547D}, where $\sigma_{v}$ is velocity dispersion. From eq. \ref{eq:2}, we can find that the $V_{rot}$ is not a direct tracer of $\Phi$ if the pressure term is non-negligible. The asymmetric drift $V_{A}$ is defined as:

\begin{equation} \label{eq:3}
V_{A}^2=V_c^2-V_{rot}^2=-\frac{R}{\rho}\frac{\partial \rho \sigma_v^2}{\partial R}
\end{equation}

, where $V_c=\sqrt{R\frac{\partial \Phi}{\partial R}}$ is the circular speed. This could explain the CO-H$\alpha$ velocity discrepancy if the $V_A$ term for H$\alpha$ source is large. That is, there is larger radial pressure supporting the ionized gas that produce H$\alpha$ emission. Under the assumption of a thin disk and isotropic velocity dispersion, $V_A$ can be expressed as\\

\begin{equation} \label{eq:4}
V_A^{2}=-R\sigma^{2}_v\frac{\partial ln(\sigma^2_v\Sigma_{obs}cos\,i)}{\partial R}
\end{equation}

Since 3DB can fit $V_{rot}$, $\sigma_{v}$ and gas surface density simultaneously, we can calculate the $V_{A}$ for CO and H$\alpha$ and determine the intrinsic circular velocity. Such a process is called asymmetric drift correction (ADC hereafter; \citealt{2017MNRAS.466.4159I}). If this scenario can fully explain the "CO-H$\alpha$ velocity discrepancy", we should expect\\

\begin{equation} \label{eq:5}
\sqrt{V_{rot,CO}^2+V_{A,CO}^2}\approx\sqrt{V_{rot,H\alpha}^2+V_{A,H\alpha}^2}
\end{equation}

\subsubsection{Notes about the asymmetric drift correction}
From Eq.\ref{eq:3} to Eq.\ref{eq:4}, one needs to assume a constant scale height and exponential distribution in the vertical direction, which is adopted in this study. Detail discussion about this assumption is in Sec \ref{sec:sec6.2}. In addition, we also assume constant filling factors, path length, and temperature for the H$\alpha$ and the X factor for the CO.\\

To obtain a smoother asymmetric drift correction, following \cite{2017MNRAS.466.4159I}, we use functional forms to describe the velocity dispersion radial profiles and the elements in the logarithm in Eq.\ref{eq:4}. The velocity dispersion is fitted by polynomials $\sigma_p(R,n_p)$ with the degree $n_p$ equal to or lower than 3. On the other hand, we fit $\sigma_{V}^2 \Sigma_{obs}cos i$ by the the function (\citealt{2002AJ....123.1316B})

\begin{equation} \label{eq:6}
f(R)=f_0(\frac{R_c}{arcsec}+1)(\frac{R_c}{arcsec}+e^{\frac{R}{R_d}})^{-1} 
\end{equation}

Where $f_0$ is the normalization factor, and $R_c$ and $R_d$ are the characteristic radii. combining Eq.\ref{eq:4} and Eq.\ref{eq:6}, we get

\begin{equation} \label{eq:7}
V_A^2=R \frac{\sigma^2_{p}(R,n_p)e^{\frac{R}{R_d}}}{R_d}(\frac{R_c}{arcsec}+e^{\frac{R}{R_d}})^{-1}
\end{equation}

No error of asymmetric drift correction is included in this study. The calculated $V_A$ for each galaxy is shown in Figure \ref{fig:ALMa_mom_rot_vdisp_1} and Figure \ref{fig:EDGE_mom_rot_vdisp_1}. As expected, H$\alpha$ tends to have higher $V_A$ due to its large velocity dispersion. In general, we did not find any radial trend of $V_A$ statistically.

\subsubsection{The CO-H$\alpha$ velocity discrepancy after asymmetric drift correction}

\begin{figure}[tbp]  
    \centering 
    \includegraphics[width=0.5\textwidth]{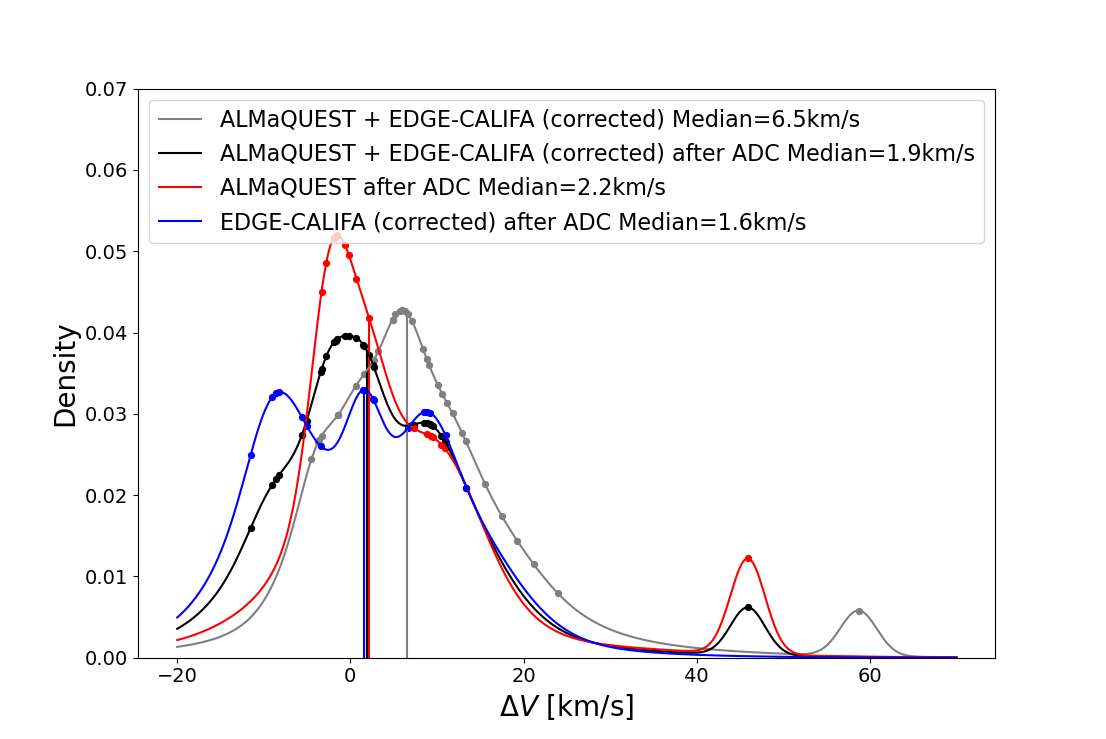}
    \caption{Kernel density distribution (KDD) of $\Delta V$ after asymmetric drift correction (ADC) for both CO and H$\alpha$ rotation curves. The gray curve is the KDD of $\Delta V$ before ADC in the ALMaQUEST+EDGE-CALIFA KSS (the gray curve in \ref{fig:kernel_density}), and the black curve is the KDD of $\Delta V$ after ADC in the ALMaQUEST+EDGE-CALIFA KSS. The red curve is the KDD of the $\Delta V$ in the ALMaQUEST sample KSS after ADC. The blue one is the KDD of the $\Delta V$ in the EDGE-CALIFA sample KSS after ADC.} 
    \label{fig:kernel_DeltaV_asy_corr}
    \end{figure}

Figure \ref{fig:kernel_DeltaV_asy_corr} shows the KDD of $\Delta V$ from both the ALMaQUEST KSS and the EDGE-CALIFA KSS after ADC, which is denoted as  $\Delta V^{\prime}$ hereafter. Compared to Figure \ref{fig:kernel_density}, we can see that $\Delta V$ is reduced in each of the samples after performing the ADC. The median value of $\Delta V^{\prime}$ becomes 1.9km/s, as opposed to the original value of $\Delta V$ ( = 6.5 km/s). This result suggests that the H$\alpha$ emission source is sustained partially by radial pressure. 

\subsection{The presence of eDIG} \label{sec:sec5.3}
Previous studies in the literature have shown that the rotation speed of extra-planar diffuse ionized gas (eDIG hereafter) decreases with increasing height above the mid-plane of the galaxy  \citep{2017ApJ...839...87B,2019ApJ...882...84L}. Such a vertical gradient in the rotation velocity could be responsible for the observed slower H$\alpha$ rotation  \citep{2018ApJ...860...92L,2020MNRAS.494.1622E,2020ARA&A..58...99S}. \cite{2018MNRAS.474.3727L} suggests that EW(H$\alpha$) is a good proxy for diagnosing the ionization mechanism. Emissions with EW(H$\alpha$)\textgreater14\AA{} trace the region where ionization is dominated by the HII region located in the mid-plane of galaxies, whereas emissions with EW(H$\alpha$)\textless3\AA{} have recently been argued to primarily come from evolved, low mass stars (i.e., HOLMES) but the origin is still not well established. While the sources of the faint ionized gas have not been well established (e.g., \citealt{2009RvMP...81..969H}), the EW(H$\alpha$) in between is likely the mixture of these two regions. Following \cite{2009RvMP...81..969H}, we use the flux-weighted average of EW(H$\alpha$) in the region where we calculate the $\Delta V$ in each galaxy to divide our galaxies into two groups based on an EW(H$\alpha$)=14\AA{} cut.  The H$\alpha$ intensity maps and the EW($\alpha$) maps are taken from the MaNGA DR15 PIPE3D value-added products (\citealt{2016RMxAA..52...21S,2016RMxAA..52..171S}). We simply assume the presence of the eDIG is negligible in the high EW(H$\alpha$) group, whereas it might be non-negligible in the low EW(H$\alpha$) group.\\

\begin{figure}[tbp] 
    \centering 
    \includegraphics[width=0.5\textwidth]{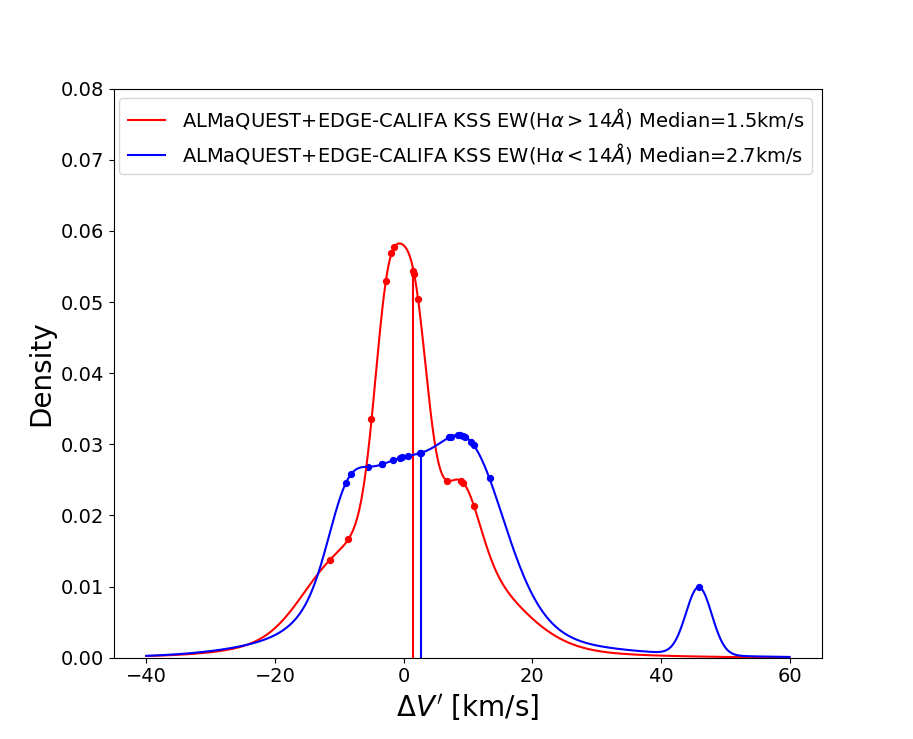}
    \caption{Kernel density distribution of $\Delta V$ after ADC for both CO and H$\alpha$ rotation curves. The whole samples are separated into two groups based on the criteria EW($H\alpha$)=14\AA{}.}
    \label{fig:deltaV_kernel_EW_sep}
    \end{figure}

We plot the $\Delta V^{\prime}$ density distribution for these two groups in Figure \ref{fig:deltaV_kernel_EW_sep}. Both groups show density distribution centered at the origin, but the high EW(H$\alpha$) group has a much smaller dispersion. We note that for the low EW(H$\alpha$) group, the ADC might not be accurate because of the following reasons: (i) the presence of eDIG ”contaminates” the H$\alpha$ velocity dispersion on the mid-plane; (ii) the assumption we made in Sec.\ref{sec:sec5.2} (e.g. thin disk, constant scale height, etc) is no longer valid. Hence, the asymmetric correction alone for these galaxies may not be accurate, which leads to a larger dispersion. Therefore, this result suggests that the eDIG is likely to play a role in the CO-H$\alpha$ velocity discrepancy, as suggested by \cite{2018ApJ...860...92L}.

\subsection{The MaNGA-CALIFA systematics} \label{sec:sec5.4}
\begin{figure}[tbp] 
    \centering 
    \includegraphics[width=0.5\textwidth]{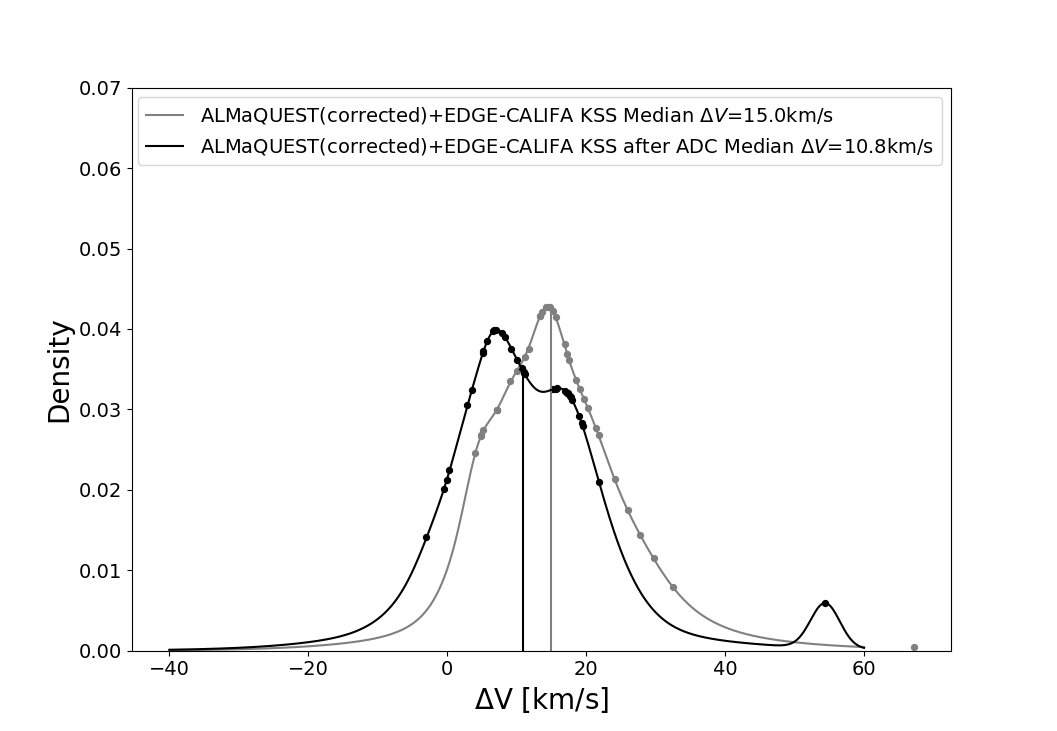}
    \caption{The kernel density distribution of $\Delta V$ and $\Delta V^{\prime}$ in ALMaQUEST+EDGE-CALIFA KSS under the scenario that the MaNGA-CALIFA systematics is corrected on the ALMaQUEST KSS.}
    \label{fig:ALMa+8.5}
    \end{figure}

The results of this work assume that the rotation curves based on the MaNGA data are accurate and the MaNGA-CALIFA systematics is atrributed to the CALIFA data. Although the exact cause of the MaNGA-CALIFA systematics is not well known, this is a reasonable choice based on the fact that the MaNGA survey has higher spectral resolution, better control of instrumental dispersion and potentially lower systematics \citep{2015ApJ...798....7B,2021AJ....161...52L}. For comparison, we discuss an opposite case, in which the offset is caused by the systematics in the MaNGA data only. We apply the correction of the MaNGA-CALIFA systematics to the ALMaQUEST KSS. In this scenario, the median value of $\Delta$V and the $\Delta V^{\prime}$ in the ALMaQUEST+EDGE-CALIFA KSS become greater, 15.0 km/s and 10.8 km/s, respectively as shown in the Figure \ref{fig:ALMa+8.5}. This result suggests that, while asymmetric drift correction again is able to further reduce the $\Delta$ V, some other mechanism(s) is needed to fully account for the CO-H$\alpha$ velocity discrepancy, which is beyond the scope of this work.


\section{Discussion}\label{sec:discussion}

\subsection{The robustness of the rotation curves and the ADC corrections} \label{Sec:Vdisp_test}
\begin{figure}[htbp] 
    \centering 
    \includegraphics[width=0.5\textwidth]{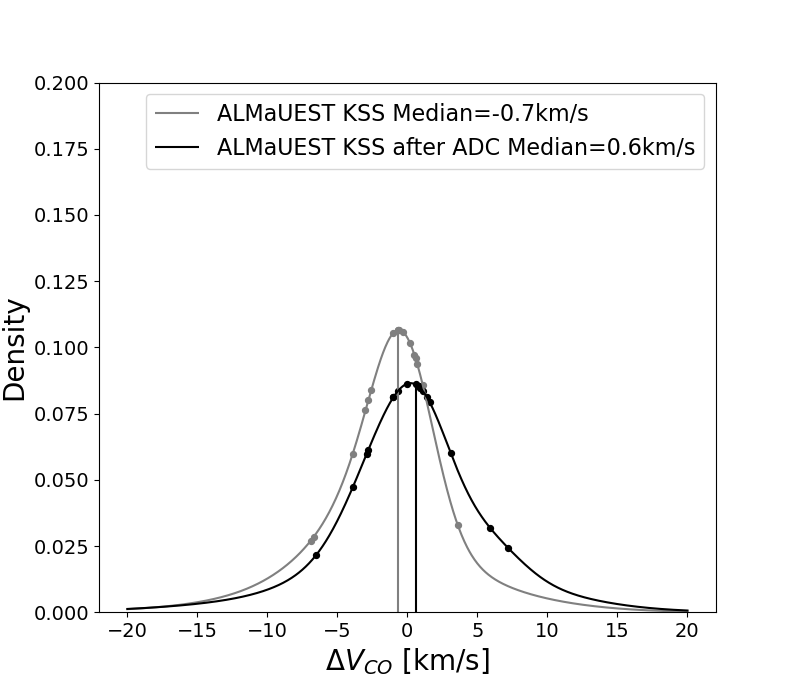}
    \caption{We compare the $V_{rot}(CO)$ fit from the original ALMaQUEST KSS datacubes, which have spectral resolution $\sim$11km/s, to the $V_{rot}(CO)$ fit from convolved ALMaQUEST KSS datacube, which has spectral resolution $\sim$116km/s. The gray curve shows the kernel density distribution of $\Delta V_{CO}$ before ADC, and the black one is for the one after ADC ($\Delta V^{\prime}_{CO}$).}
    \label{fig:CO_conv_test}
    \end{figure}

The spectral resolutions of both MaNGA IFS ($\sigma_{inst} \approx $70 km/s) and EDGE-CALIFA survey ($\sigma_{inst} \approx$116 km/s) of H$\alpha$ are in general higher than the intrinsic astrophysical dispersion $\sigma_{H\alpha}$. Hence, it is crucial to test if the relatively low spectral resolution of H$\alpha$ influence the rotation curves and the ADC corrections. Hence, We convolve our ALMaQUEST KSS CO datacubes, which have original spectral resolution $\sigma_{inst} \approx$ 11 km/s, to the resolution of the CALIFA H$\alpha$ observation ($\sigma_{inst} \approx$ 116 km/s) with a Gaussian kernel and use 3DB to fit the CO datacube again. After this process, following the previous procedure, we compare the rotation curves derived from the original CO data to the rotation curves derived from the convolved CO data (using the same method described in Sec.\ref{sec:sec5.1}, denoted as $\Delta V_{CO}$). The comparison of the rotation curves after ADC correction is denoted as $\Delta V^{\prime}_{CO}$. The result is shown in Figure \ref{fig:CO_conv_test}. As we can see, both the $\Delta V_{CO}$ and the $\Delta V^{\prime}_{CO}$ density distribution show a Gaussian-like distribution centered at zero. By combining this result and the fact that the velocity dispersion of H$\alpha$ is usually higher than that of CO, we conclude that: (i) The systematic velocity discrepancy between CO and H$\alpha$ we found is not sensitive to their different spectral resolutions within the range considered in this work; (ii) Statistically, the ADC applied to the rotation curves presented in this work are reliable. 

\subsection{Thin disk assumption}  \label{sec:sec6.2}
\begin{figure}[htbp] 
    \centering 
    \includegraphics[width=0.5\textwidth]{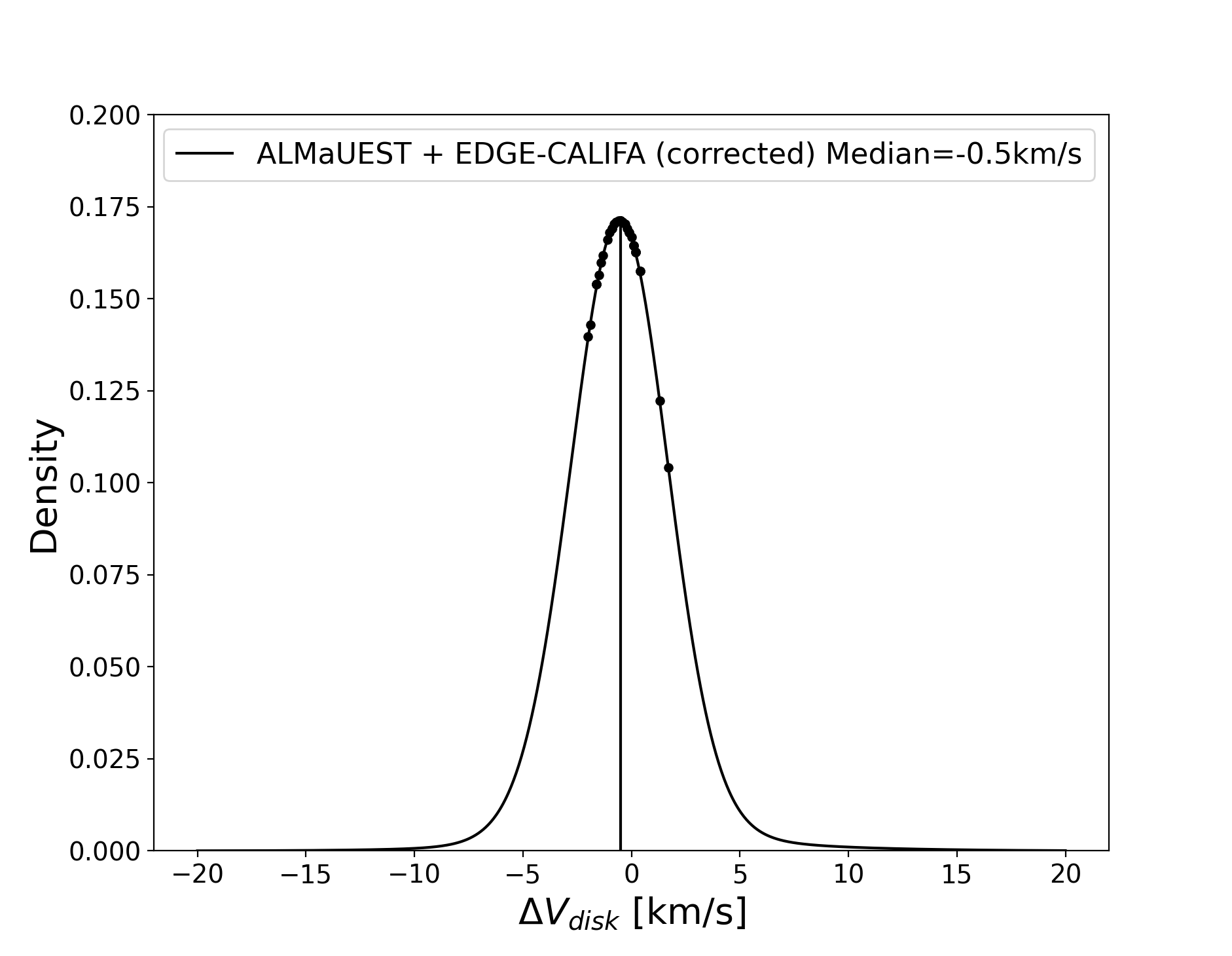}
    \caption{The $\Delta V$ distribution between the $V_{rot,z=100pc}$ and the $V_{rot,z=1000pc}$.}
    \label{fig:thick_disk_test1}
    \end{figure}

The discussion in this work so far relies on the thin disk assumption. With the presence of eDIG, the thin disk assumption is no longer valid. Hence, it is crucial to investigate whether thin disk assumption influences our result or not. The most straightforward test to do is use 3DB to fit our galaxies again with setting a larger scale height. We re-fit the ALMaQUEST KSS and the EDGE-CALIFA KSS again with a scale height of 1 kpc, and then compare the fit result to the ones with a scale height of 100 pc. The comparison is shown in Figure \ref{fig:thick_disk_test1}. As we can see, there is little impact on the rotation velocity. The same result is found on the velocity dispersion.\\

\begin{figure}[htbp] 
    \centering 
    \includegraphics[width=0.5\textwidth]{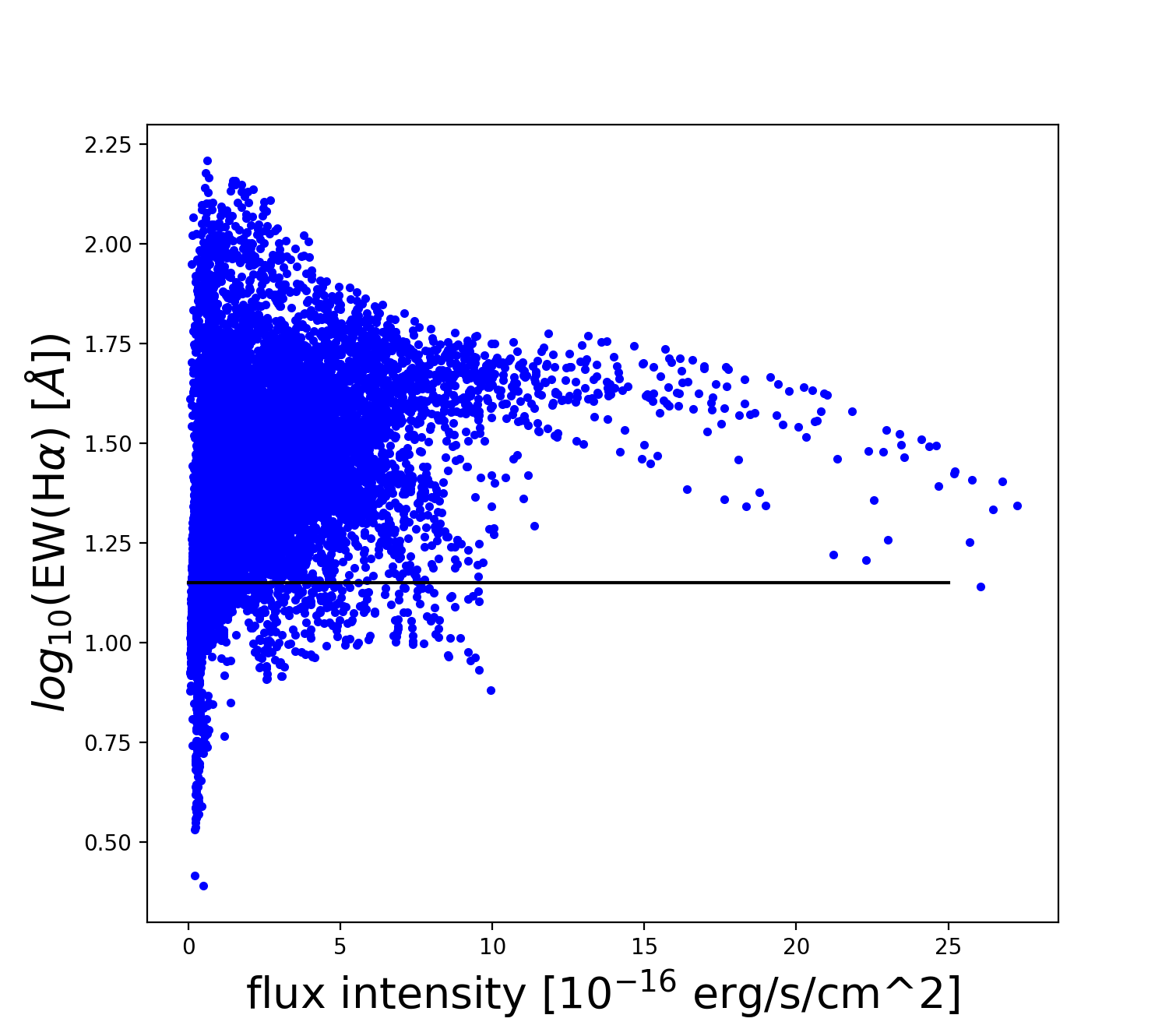}
    \caption{The scatter plots of spaxels which lie in the regions we calculate $\Delta V$ in the high EW group. The black horizontal line is EW=14\AA{}.}
    \label{fig:thick_disk_test2}
    \end{figure}

Figure \ref{fig:thick_disk_test2} show the scatter plots of spaxels which lie in the region where we calculate $\Delta V$ in the large $EW(H\alpha)$ group. As we can see, most of the spaxels (90\%) in these regions have $EW(H{\alpha})$ larger than 14 \AA{}, which are usually HII regions. Hence, the presence of DIG has little impact on the 3DB fit results for the high EW group galaxies. Therefore, we conclude that the thin disk model and a constant scale height are reasonable for these galaxies and the main conclusion of this work still holds.


\subsection{Dynamical mass measurement}
From the result presented above, one can find that the mass model based on ionized gas would potentially be biased low. For a galaxy with an intrinsic circular velocity of 200 km/s (the median value in our sample), the $\Delta V=6.5$ km/s (the median value of $\Delta V$ in our sample) would underestimate the total dynamical mass by $\sim$6\%. In this study, we find that by applying ADC, the discrepancy between Ha and CO rotation curve can be further reduced to 1.9 km/s, corresponding to ~2\% underestimation in the dynamical mass measurement based on the Ha rotation curve. In the future, we will compare dynamical mass derived from the H$\alpha$ rotation curves in our galaxies with other's studies, such as \cite{2018NatAs...2..233Z} and \cite{2020ApJ...900..109A}.

\section{Conclusion}\label{sec:conclusion}
By combining the ALMaQUEST and the EDGE-CALIFA surveys, we analyze and compare the CO and H$\alpha$ rotation curves in 34 rotational-dominated galaxies. Before combining the two datasets, we first compare the H$\alpha$ rotation curves in 8 MaNGA-CALIFA overlapping galaxies. We find systematics between the H$\alpha$ rotation curves measured from the MaNGA data and that from the CALIFA data, where the CALIFA rotation curves tend to have slower rotation velocities by 8.5 km/s. We compensate the systematics by adding the 8.5 km/s to the CALIFA dataset, and the conclusion is based on this premise. Our principal conclusions are:

\begin{itemize}
    \item 56 $\%$ of our galaxies show smaller H$\alpha$ rotation velocity (10/17 in the ALMaQUEST KSS, and 9/17 in the EDGE-CALIFA KSS). The median value of the CO-H$\alpha$ rotation velocity difference is 6.5 km/s (5.7 km/s for the ALMaQUEST samples, and 6.7 km/s for the EDGE-CALIFA samples). The remaining 44\% of the sample shows consistency between CO and H$\alpha$ rotations. For a galaxy with circular velocity 200km/s (the median value in our sample), $\Delta V=$6.5 km/s would lead to $\sim$6\% underestimation in the dynamical mass.
    
    \item The magnitude of velocity differences between CO and H$\alpha$ rotation velocity, $\Delta V$, correlates with the difference between CO and H$\alpha$ velocity dispersion, which suggests the gas radial pressure gradient plays a key role to explain the CO-H$\alpha$ velocity discrepancy. Under the assumption that the turbulence pressure dominates the radial pressure on the H$\alpha$ emitting gas, we apply the asymmetric drift correction (ADC) to both CO and H$\alpha$ rotation curves. After ADC, the median value of the CO-H$\alpha$ rotation velocity difference reduces to 1.9 km/s (2.2 km/s for ALMaQUEST KSS, and 1.6 km/s for EDGE-CALIFA KSS), which corresponds to $\sim$2\% underestimation of dynamical mass based on the Ha measurement in the case of the circular velocity being 200 km/s.
    
    \item We divided our sample into two groups based on the EW(H$\alpha$)=14\AA{} cut. The ionization in the high EW group is dominated by the HII regions. For the low EW group, the contribution of extra-planar diffuse ionized gas to the H$\alpha$ emission is non-negligible. The median value of $\Delta V$ after ADC is 1.5 km/s in the high EW group, which is 2.7 km/s in the low EW group. Both groups show density distribution centered at the origin, but the low EW(H$\alpha$) group has a much larger dispersion, which could be due to the presence of extra-planar diffuse ionized gas (eDIG).

Our result suggests both radial pressure gradient and the presence of extra-planar diffuse ionized gas are responsible for the velocity difference between CO and H$\alpha$ rotations. This result suggests it is possible to recover the circular velocity from the H$\alpha$ rotation curves if properly taking these two factors into accounts.

\end{itemize}

\appendix
\renewcommand\thefigure{\thesection.\arabic{figure}}
\setcounter{figure}{0}
\section{Star formation rate surface density}
\label{appen:SFR_surface_density}
To calculate the $\Sigma_{SFR}$ within the region where we determine $\Delta V$, for each galaxy, we first pick up the spaxel in which the signal-to-noise ratio (S/N)$>$3 in the emission lines $H\alpha$, $H\beta$, [OIII]$\lambda$5007 and [SII]$\lambda \lambda 6717,31$. To take into account the effect of dust extinction on SFR, we use the Balmer decrement to correct the H$\alpha$ flux based on the equation which assumes a \cite{1989ApJ...345..245C} extinction curve

\begin{equation}\label{eq:A1}
A_{H_\alpha}=5.86\,log(\frac{F_{H_\alpha}}{2.86F_{H_\beta}})
\end{equation}

Where $F_{H_\alpha}$ and $F_{H_\beta}$ are the fluxes of H$\alpha$ and H$\beta$ emission line, respectively. Then, we calculate the SFR spaxel-by-spaxel based on the the equation \citep{2002MNRAS.332..283R}

\begin{equation}\label{eq:A2}
SFR=7.9 \times 10^{-42} F_{H_\alpha} 10^{\frac{A_{H_\alpha}}{2.5}}
\end{equation}

, which assumes a Salpeter Initial Mass Function (IMF). When calculating the $\Sigma_{SFR}$, only the star forming spaxels classified with the [SII] BPT diagnostic \citep{2001ApJ...556..121K, 2006MNRAS.372..961K} are included. Finally, we sum up the SFR spaxels within the region where we determine $\Delta V$ and divided it by the deprojected area the inclination correction applied.

\section{The test on the MaNGA-CALIFA overlapping galaxies}\label{appen:MC_test}

This section aimed at investigating the systematic between the MaNGA and the CALIFA H$\alpha$ data. We match the MaNGA and the CALIFA galaxies \citep{2016RMxAA..52...21S,2016RMxAA..52..171S}, and 37 overlapping galaxies are found. To make the result applicable to the galaxies in our study, we apply the same selection criteria described in Sec \ref{sec:4.2}, and finally, 8 galaxies are included in this test.\\

We use 3DB to fit the H$\alpha$ rotation curves for these overlapping galaxies using both the MaNGA and the CALIFA datacubes. We follow the strategy described in Sec \ref{sec:analysis}. For each galaxy, we use the same inclination and position for the MaNGA and the CALIFA fitting. The inclinations and the position angles are based on the MaNGA best fit.\\

\subsection{The comparison of the rotation curves}

\begin{figure}[htbp] 
    \centering 
    \includegraphics[width=0.5\textwidth]{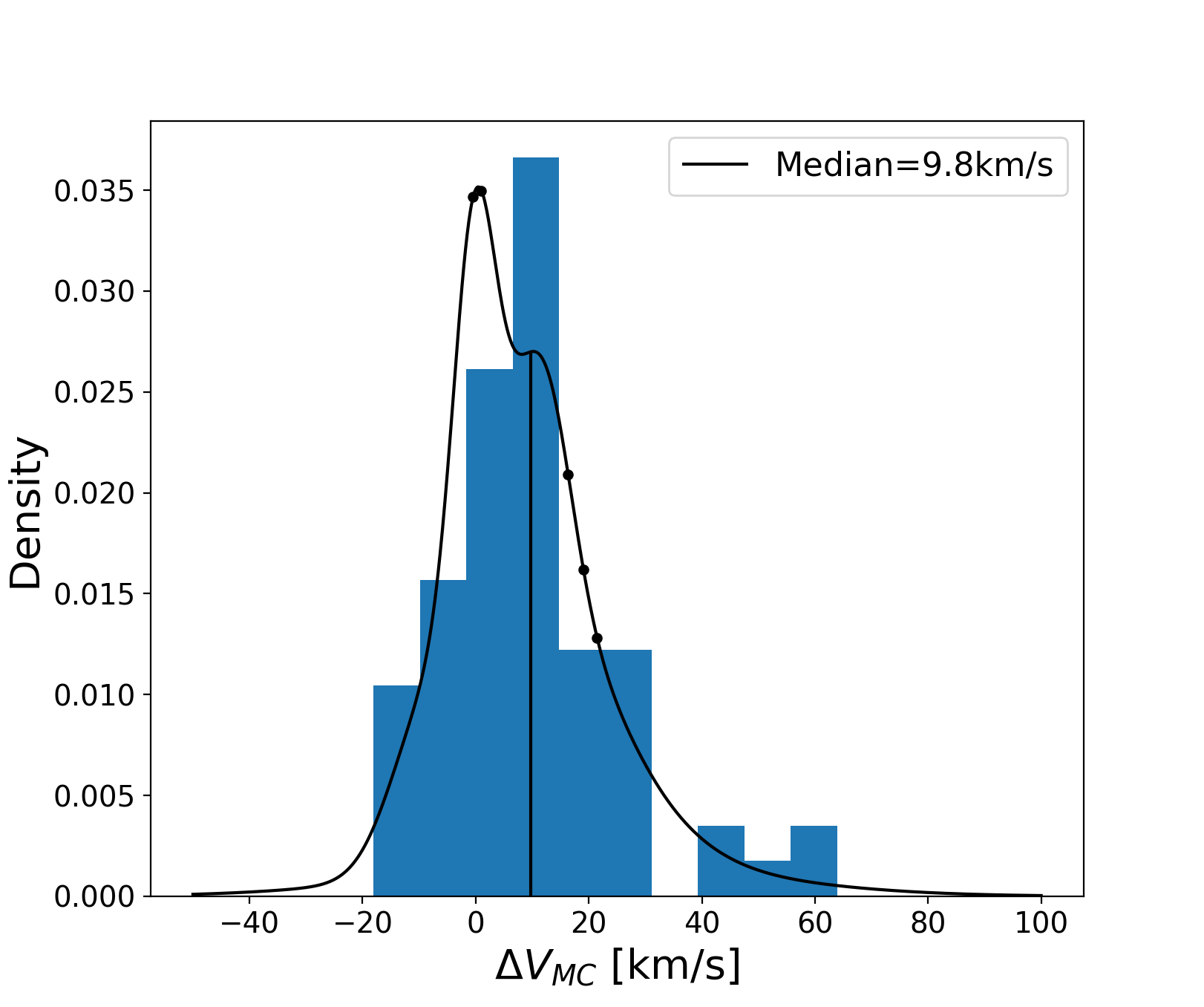}
    \caption{The $\Delta V_{MC}$ distribution. The blue bins are the histogram of the velocity difference of radial bins in all galaxies. This result shows that the CALIFA fit result tends to have slower rotation velocity by 9.8 km/s.}
    \label{fig:MC_test1}
    \end{figure}

After comparing the MaNGA and the CALIFA fitting result in these overlapping galaxies, we found a systematic in the rotation velocity. The best fit of the CALIDA data tends to have slower rotation velocity by 9.8 km/s, as shown in Figure \ref{fig:MC_test1}. We adopt the definition of $\Delta V$ and use it to quantify the difference between the MaNGA and the CALIFA rotation curves fit results in these overlapping galaxies, denoted as $\Delta V_{MC}$.
To test the robustness of $\Delta V_{MC}$, we use bootstrap to re-sampling the velocity differences of the radial bins in all galaxies a number of times and calculate the median value of $\Delta V_{MC}$. Finally, the systematics of 8.5 km/s with a dispersion of 1.3 km/s is found in the median value of $\Delta V_{MC}$. This result suggests there is systematics between the MaNGA fit result and the CALIFA fit result based on our method. Given that MaNGA data has the better spectral resolution, we correct the $\Delta V$s (i.e. the variance weighted mean of the velocity differences $V_{rot,CO}-V_{rot,H\alpha}$, as defined in Sec \ref{sec:sec5.1}) in the EDGE-CALIFA KSS vaule by subtracting 8.5 km/s.

\subsection{The comparison of the velocity dispersion radial profiles}
The robustness of the velocity dispersion radial profiles are crucial for calculating the reliable asymmetric drift correction. Compared to H$_{\alpha}$, the measurement of the CO velocity dispersion should be more robust due to its higher spectral resolution. On the other hand, \cite{2021AJ....161...52L} found that the line spread function (LSF) of MaNGA H$\alpha$ lines can be described well by a Gaussian, and the reliability of the astrophysical velocity dispersion of 20 km/s could be achieved when the S/N is sufficiently high enough. The low spectral resolution of the CALIFA data, however, raises the concern of the reliability of the H$\alpha$ velocity dispersion measurement in the EDGE-CALIFA KSS. Under the condition that the spectral resolution is much lower than the astrophysical velocity dispersion, the LSF must be well known to get the reliable velocity dispersion. In this study, we adopt the instrumental dispersion of all CALIFA H$\alpha$ data to be 116 km/s \citep{2016A&A...594A..36S}. To validate this choice, we compare the H$\alpha$ velocity dispersion radial profiles in these 8 overlapping galaxies measured between the MaNGA data and the CALIFA data. \\

\begin{figure}[htbp] 
    \centering 
    \includegraphics[width=0.5\textwidth]{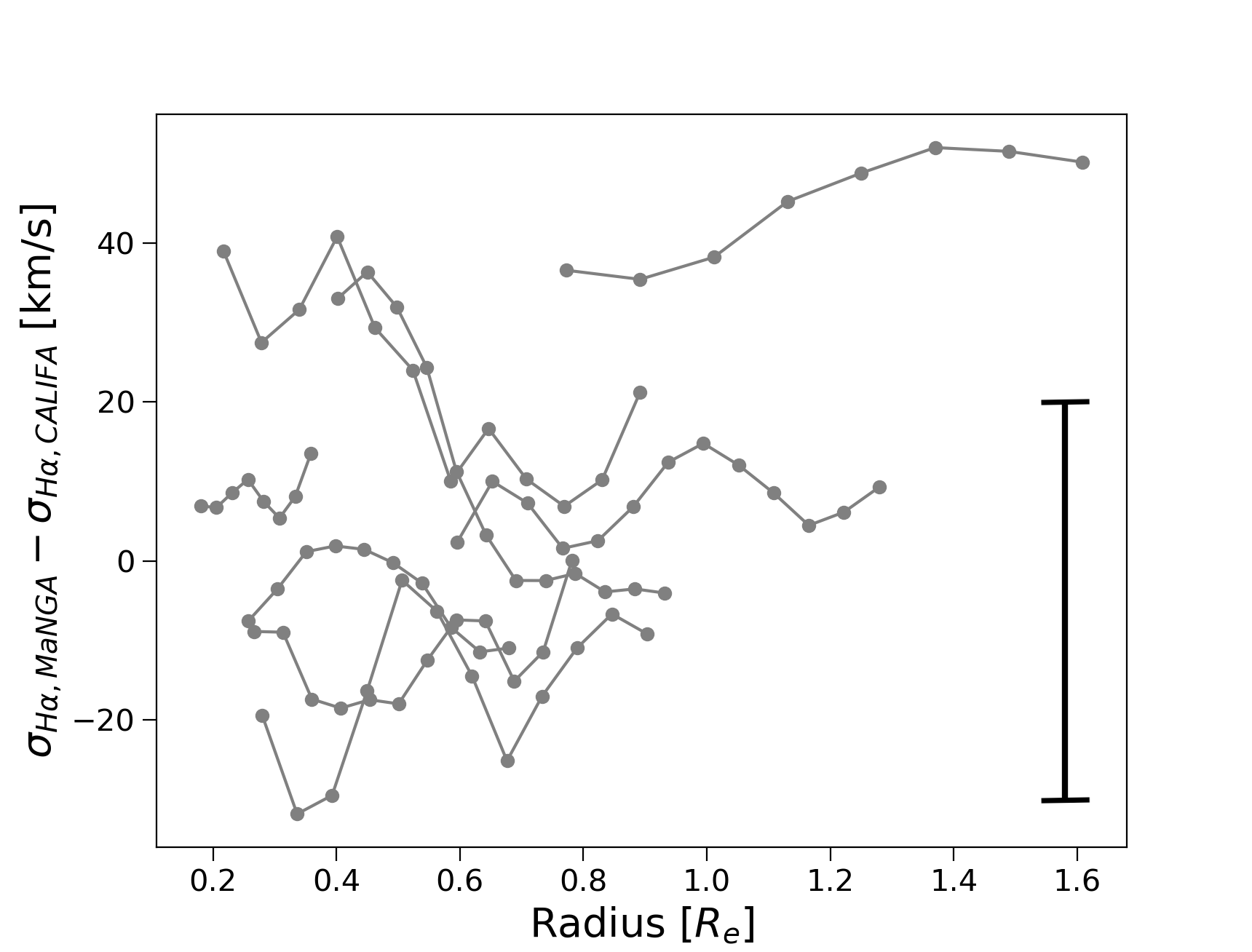}
    \caption{The vertical axis is the difference in the velocity dispersion between the MaNGA and CALIFA data for the 8 MaNGA-CALIFA overlapping galaxies. The horizontal axis is the radius in effective radius. Each gray line presents one galaxy, and the vertical black line presents the typical uncertainty.}
    \label{fig:MC_Halpha_delta_Vdisp}
    \end{figure}

In Fig \ref{fig:MC_Halpha_delta_Vdisp}, the gray lines are the radial profile of the difference in the velocity dispersion between the MaNGA measurement and the CALIFA measurement for the 8 overlapping galaxies. The black vertical line is the typical uncertainty. As one can see, there is no apparent systematics in the velocity dispersion between these two measurements, suggesting that the typical instrumental dispersion we use for the CALIFA galaxies should be representative. On the other hand, the scatter of the velocity dispersion offset simply means the asymmetric drift correction measurement is subject to the uncertainty in the individual velocity dispersion measurement, which is not considered in this study. 


\section{Other plots}\label{appen:other}

    \begin{figure}[bpt] 
    \centering 
    \includegraphics[width=1\textwidth]{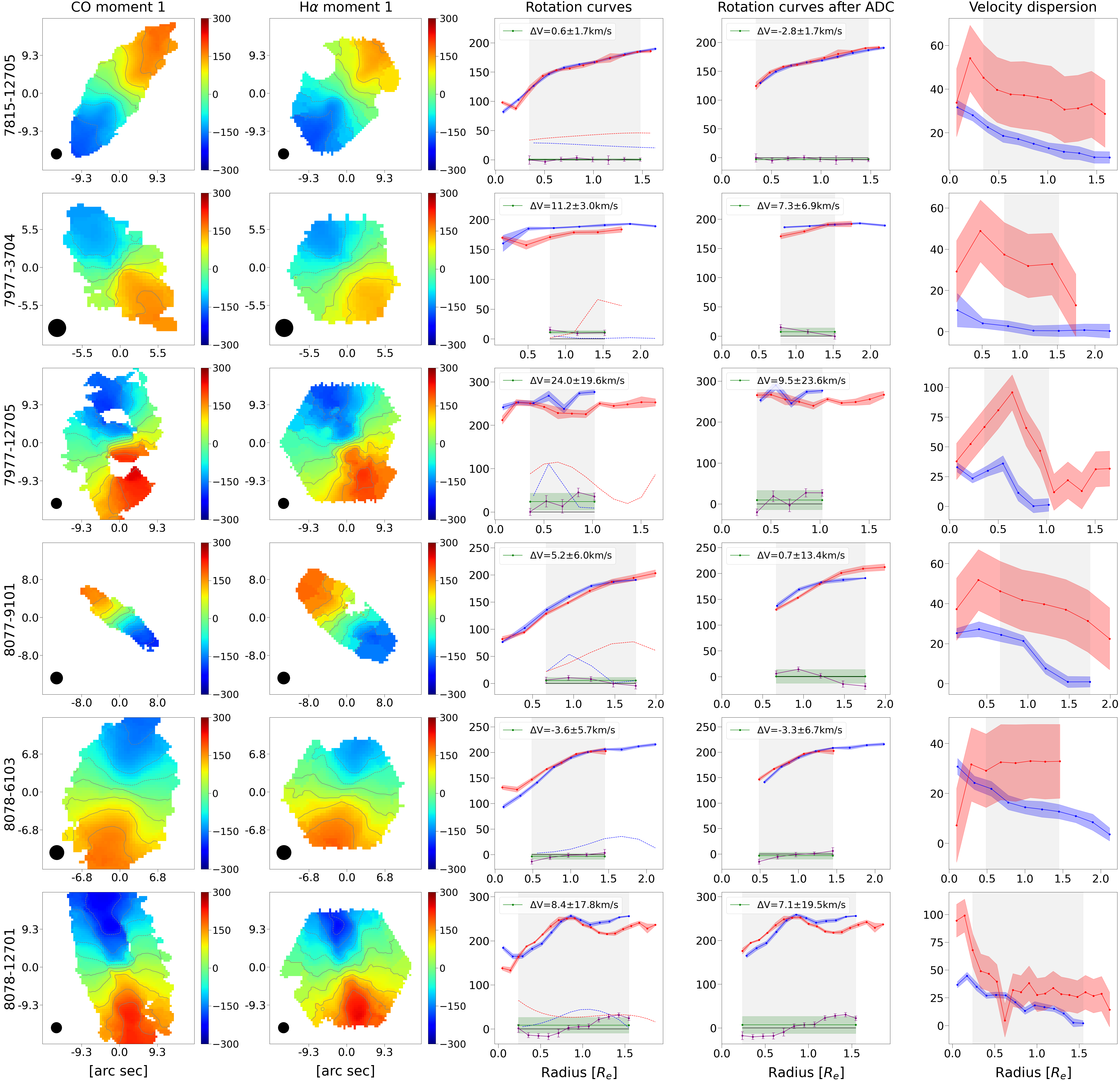}
    \caption{The first column shows the CO velocity maps for the ALMaQUEST KSS. The color in these maps varies between $\pm$300 km/s with negative velocity in blue. The interval of the contours is 50km/s. Both the axis values show the offset in arcsec from the center of the image. The black circles at the bottom left are the typical resolution of ALMaQUEST KSS data, which is \arcsec{2}{5}. The second column shows the H$\alpha$ velocity maps for the ALMaQUEST KSS with the features are identical to the first column. The third column shows the CO and H$\alpha$ rotation curves. The x-axis is the galactocentric radius in the unit of effective radius ($R_e$), and the y-axis is the velocity in km/s. Blue ones are CO rotation curves and red ones are H$\alpha$ rotation curves. The purple lines connect the velocity differences between $V_{rot}$(CO) and $V_{rot}$(H$\alpha$) at each radial bin. $\Delta$V, which is the variance weighted mean of the velocity differences, is calculated within the gray area. The error of $\Delta$V is the standard deviation of the velocity differences in sample points. The green line shows the value of $\Delta$V. The shadow areas present the errors of each quantity. The blue and red dash lines present the $V_A$ radial profiles of CO and H$\alpha$, respectively. The fourth column shows the CO and H$\alpha$ rotation curves after asymmetric drift corrections (ADC). The features in the fourth column are identical to the third column. The fifth column shows the CO and H$\alpha$ velocity dispersion radial profile. The x-axis is the galactocentric radius in the unit of $R_e$ and the y-axis is the velocity dispersion in km/s. The CO rotation curves are shown in blue and the H$\alpha$ rotation curves are shown in red.}
    \label{fig:ALMa_mom_rot_vdisp_1}
    \end{figure}
\addtocounter{figure}{-1}
    \begin{figure}[tbp] 
    \centering 
    \includegraphics[width=1\textwidth]{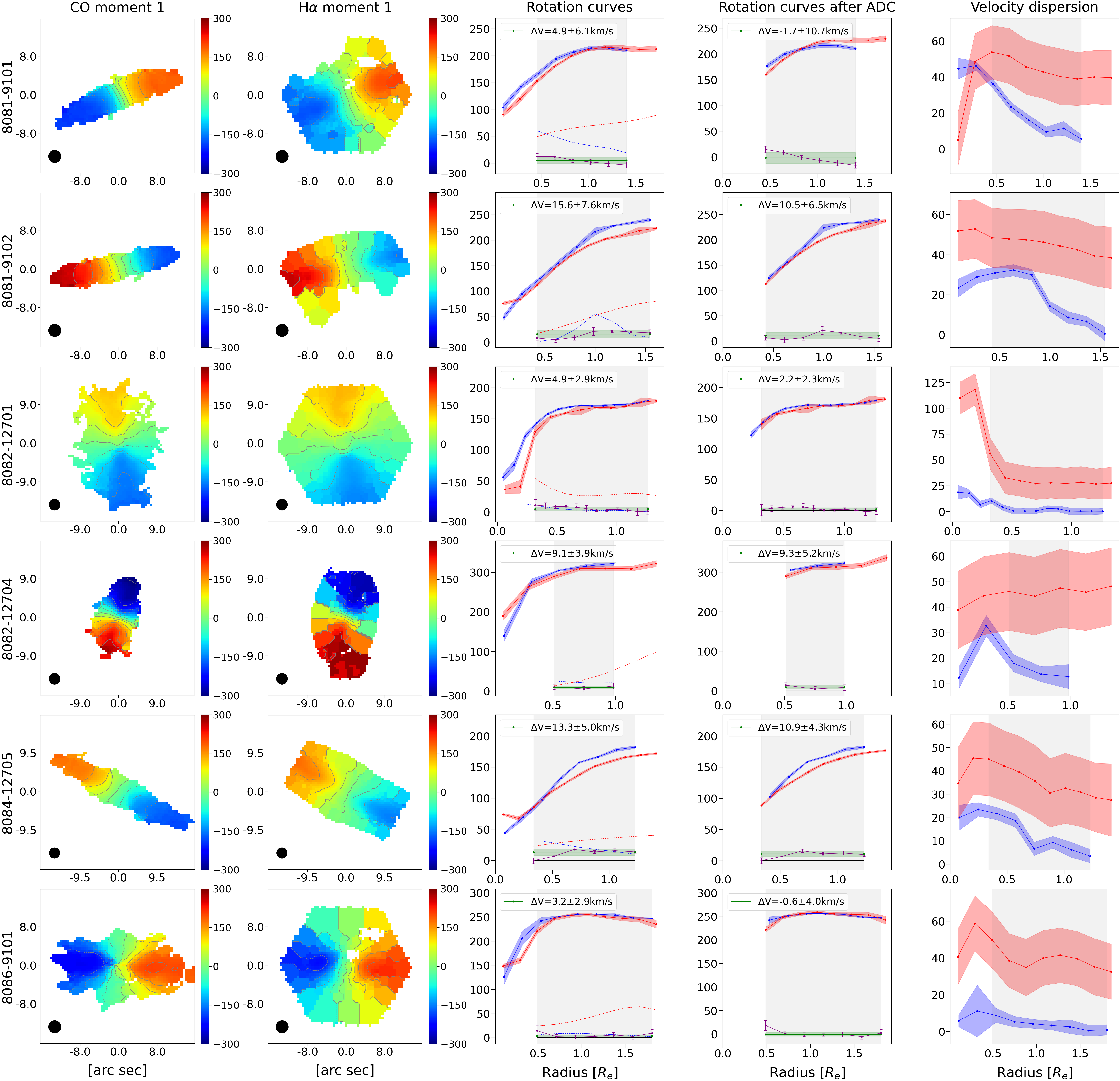}
    \caption{(Continued.)}
    \label{fig:ALMa_mom_rot_vdisp_2}
    \end{figure}
\addtocounter{figure}{-1}
    \begin{figure}[bpt] 
    \centering 
    \includegraphics[width=1\textwidth]{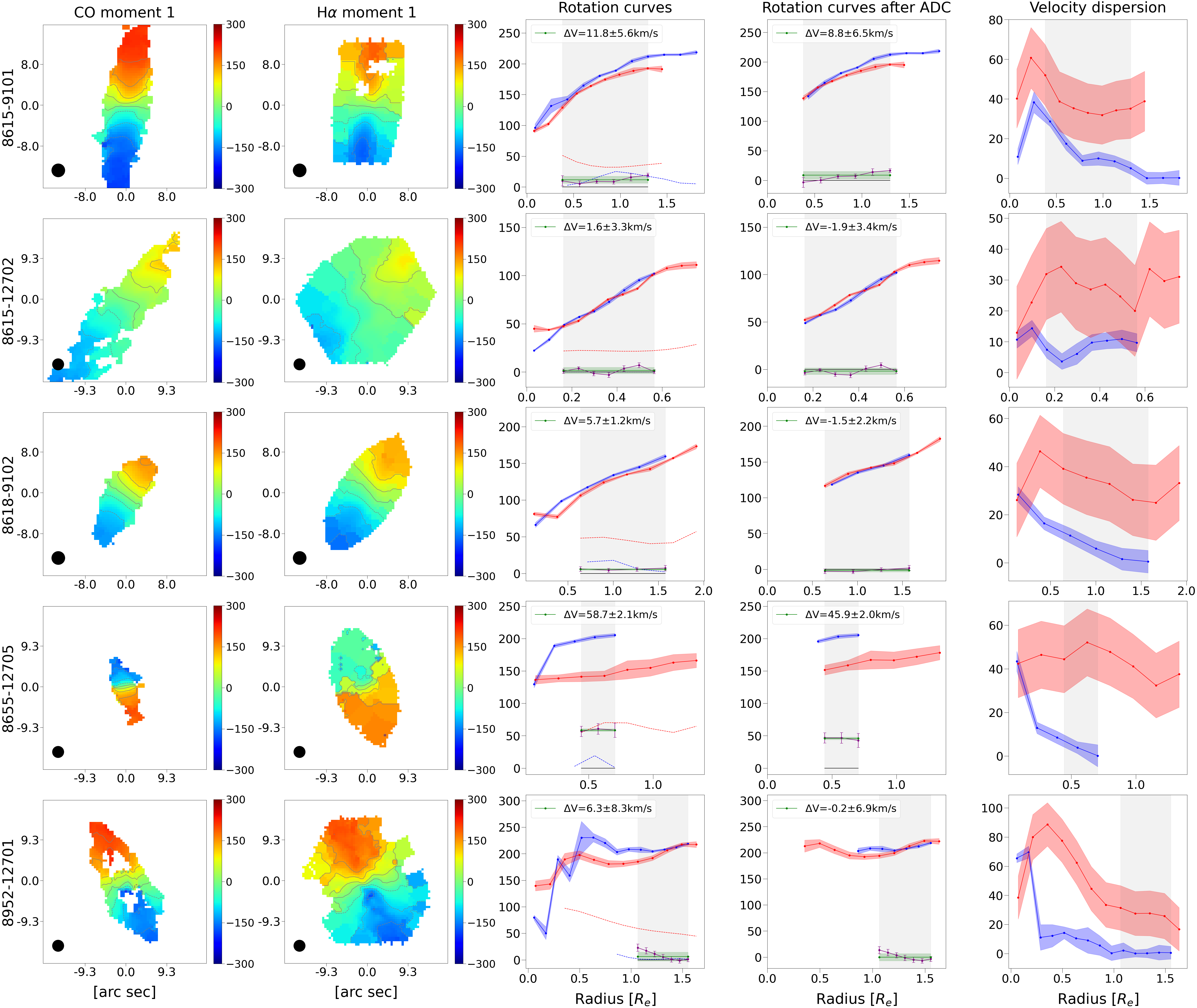}
    \caption{(Continued.)}
    \label{fig:ALMa_mom_rot_vdisp_3}
    \end{figure}
    \begin{figure}[tbp] 
    \centering 
    \includegraphics[width=0.98\textwidth]{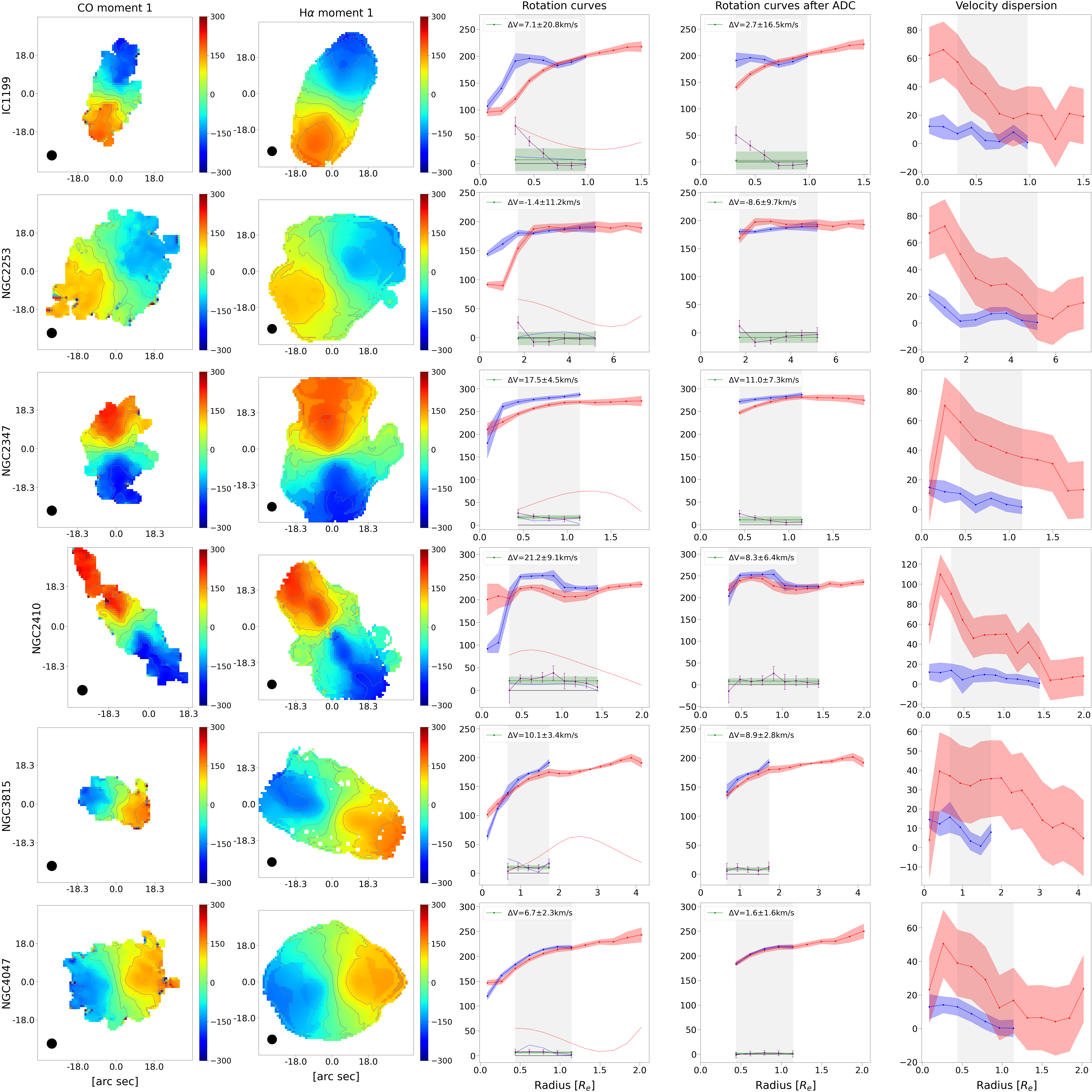}
    \caption{As Fig \ref{fig:ALMa_mom_rot_vdisp_1} but for EDGE-CALIFA KSS galaxies. The black circles at the bottom left in the first two columns are the typical resolution of EDGE-CALIFA KSS data, which is 4.5". We note that the H$\alpha$ rotation curves (and hence the $\Delta V$) in the third and the fourth columns have been corrected for the MaNGA-CALIFA systematics.}
    \label{fig:EDGE_mom_rot_vdisp_1}
    \end{figure}
\addtocounter{figure}{-1}
    \begin{figure}[tbp] 
    \centering 
    \includegraphics[width=1\textwidth]{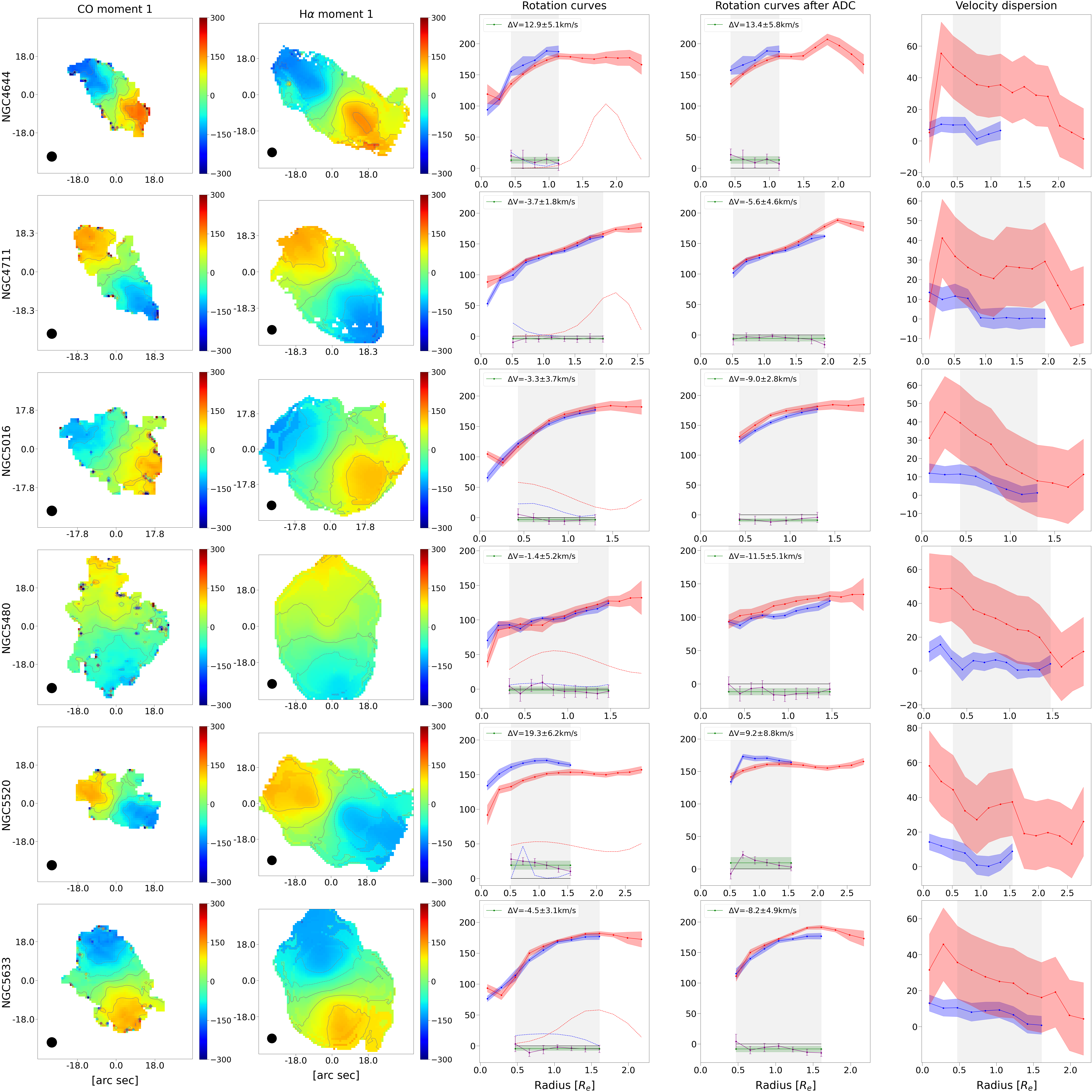}
    \caption{(Continued.)}
    \label{fig:EDGE_mom_rot_vdisp_2}
    \end{figure}
\addtocounter{figure}{-1}
    \begin{figure}[tbp] 
    \centering 
    \includegraphics[width=1\textwidth]{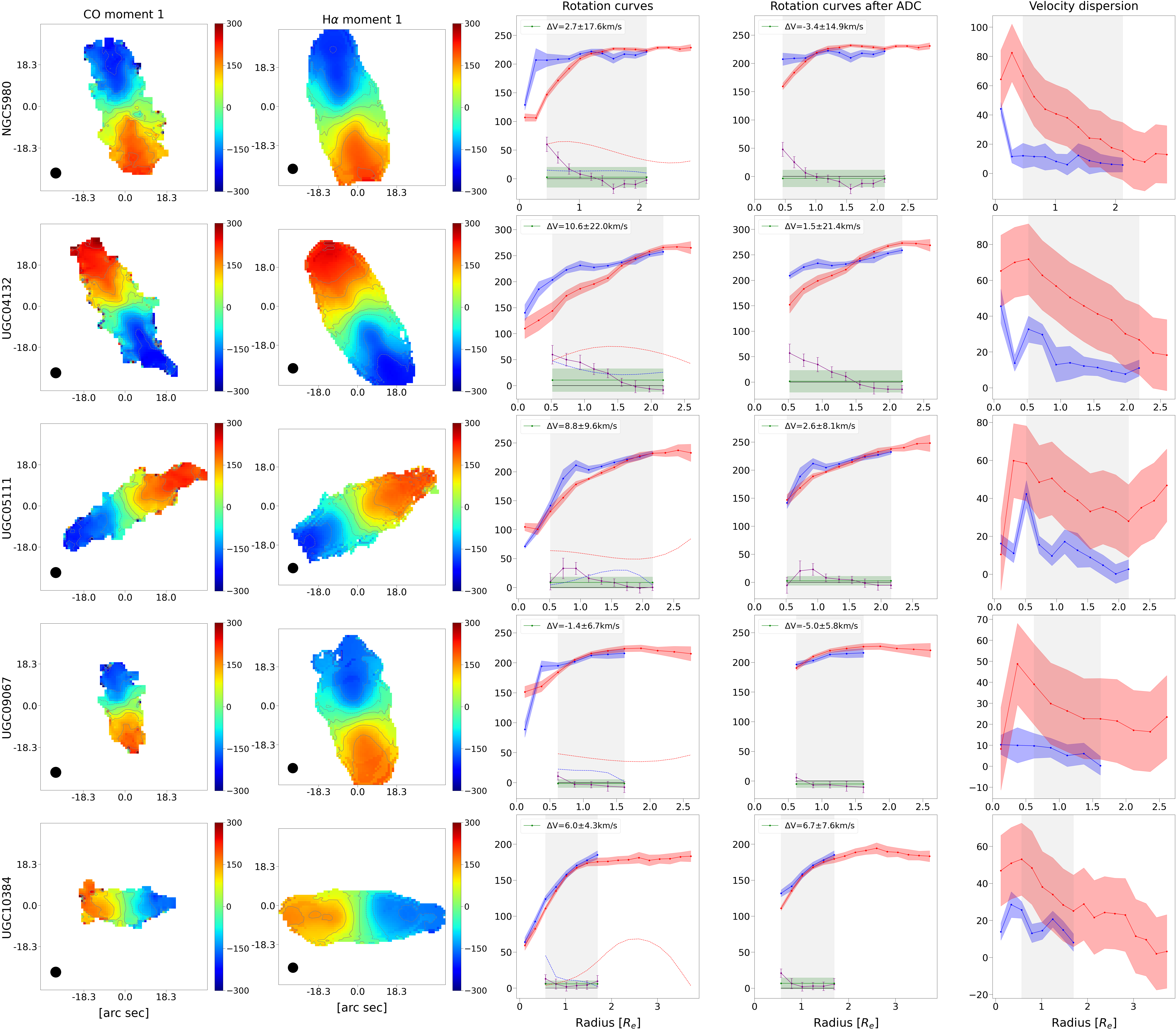}
    \caption{(Continued.)}
    \label{fig:EDGE_mom_rot_vdisp_3}
    \end{figure}
    \begin{figure}[bpt] 
    \centering 
    \includegraphics[width=0.9\textwidth]{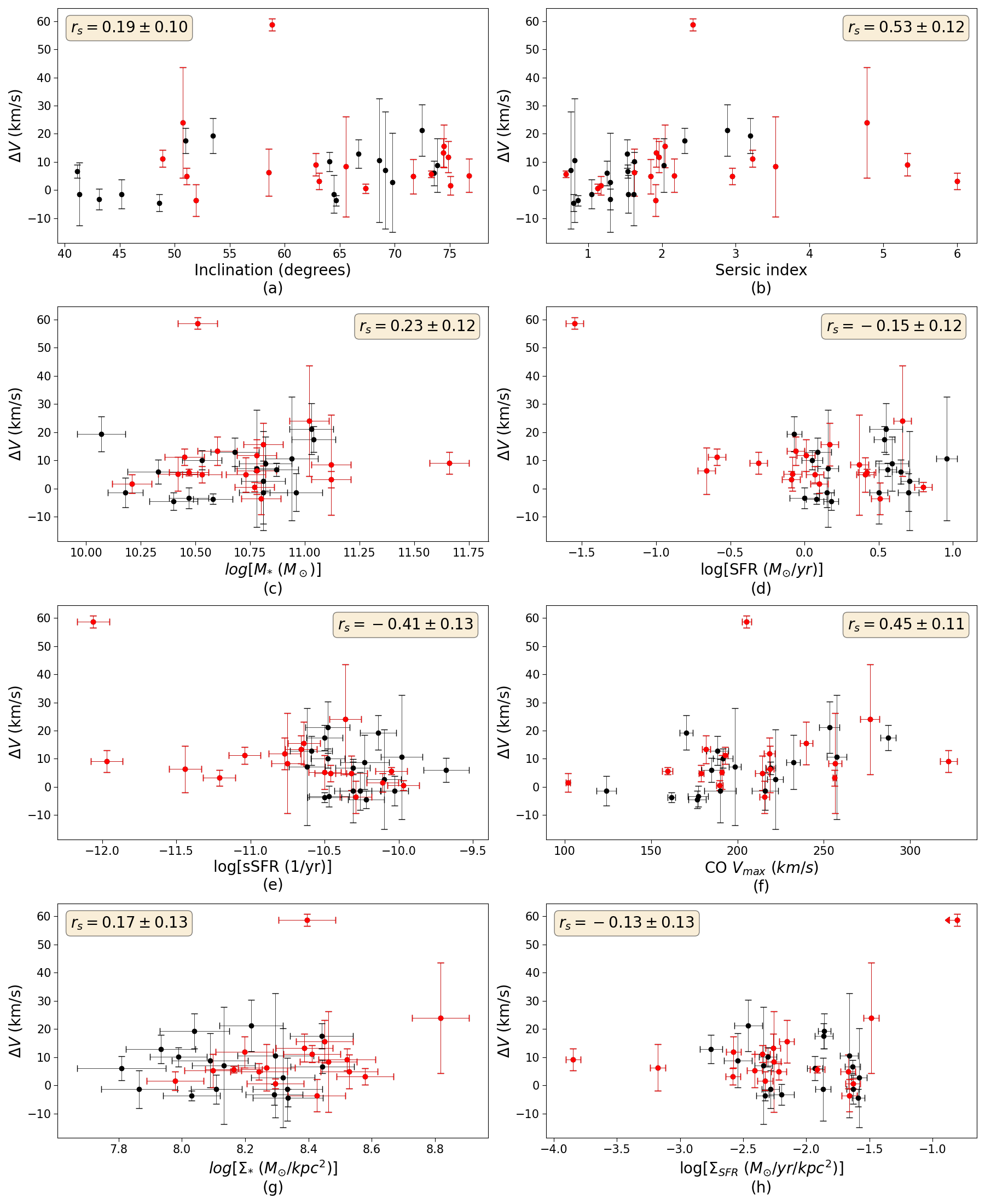}
    \caption{No apparent trend is found between $\Delta$V and (a)Inclination, (b)Sersic index, (c) stellar mass, (d) SFR, (e) sSFR, (f)CO $V_{max}$, (e) $\Sigma_{*}$ and (h)$\Sigma_{SFR}$. The red points are ALMaQUEST KSS, and the black points are the EDGE-CALIFA KSS. The Sersic index comes from SDSS nsa catalog. For the ALMaQUEST KSS galaxies, the stellar mass and SFR are listed in \cite{2020ApJ...903..145L}. For the EDGE-CALIFA KSS galaxies, the stellar mass and SFR comes from \cite{2017ApJ...846..159B}. The $\Sigma_{*}$ and $\Sigma_{SFR}$ in this figure are calculated within the region where $\Delta V$ be calculated. Check Appendix \ref{appen:SFR_surface_density} for more details.}
    \label{fig:deltaV_other_parameters}
    \end{figure}
%


\bibliography{sample63}{}

\begin{thebibliography}{}
\expandafter\ifx\csname natexlab\endcsname\relax\def\natexlab#1{#1}\fi
\providecommand{\url}[1]{\href{#1}{#1}}
\providecommand{\dodoi}[1]{doi:~\href{http://doi.org/#1}{\nolinkurl{#1}}}
\providecommand{\doeprint}[1]{\href{http://ascl.net/#1}{\nolinkurl{http://ascl.net/#1}}}
\providecommand{\doarXiv}[1]{\href{https://arxiv.org/abs/#1}{\nolinkurl{https://arxiv.org/abs/#1}}}

\bibitem[{{Aquino-Ort{\'\i}z} {et~al.}(2020){Aquino-Ort{\'\i}z}, {S{\'a}nchez},
  {Valenzuela}, {Hern{\'a}ndez-Toledo}, {Jin}, {Zhu}, {van de Ven},
  {Barrera-Ballesteros}, {Avila-Reese}, {Rodr{\'\i}guez-Puebla}, \&
  {Tissera}}]{2020ApJ...900..109A}
{Aquino-Ort{\'\i}z}, E., {S{\'a}nchez}, S.~F., {Valenzuela}, O., {et~al.} 2020,
  \apj, 900, 109, \dodoi{10.3847/1538-4357/aba94e}

\bibitem[{{Begeman}(1989)}]{1989A&A...223...47B}
{Begeman}, K.~G. 1989, \aap, 223, 47

\bibitem[{{Bizyaev} {et~al.}(2017){Bizyaev}, {Walterbos}, {Yoachim}, {Riffel},
  {Fern{\'a}ndez-Trincado}, {Pan}, {Diamond-Stanic}, {Jones}, {Thomas},
  {Cleary}, \& {Brinkmann}}]{2017ApJ...839...87B}
{Bizyaev}, D., {Walterbos}, R.~A.~M., {Yoachim}, P., {et~al.} 2017, \apj, 839,
  87, \dodoi{10.3847/1538-4357/aa6979}

\bibitem[{{Blanton} {et~al.}(2011){Blanton}, {Kazin}, {Muna}, {Weaver}, \&
  {Price-Whelan}}]{2011AJ....142...31B}
{Blanton}, M.~R., {Kazin}, E., {Muna}, D., {Weaver}, B.~A., \& {Price-Whelan},
  A. 2011, \aj, 142, 31, \dodoi{10.1088/0004-6256/142/1/31}

\bibitem[{{Bolatto} {et~al.}(2017){Bolatto}, {Wong}, {Utomo}, {Blitz}, {Vogel},
  {S{\'a}nchez}, {Barrera-Ballesteros}, {Cao}, {Colombo}, {Dannerbauer},
  {Garc{\'\i}a-Benito}, {Herrera-Camus}, {Husemann}, {Kalinova}, {Leroy},
  {Leung}, {Levy}, {Mast}, {Ostriker}, {Rosolowsky}, {Sandstrom}, {Teuben},
  {van de Ven}, \& {Walter}}]{2017ApJ...846..159B}
{Bolatto}, A.~D., {Wong}, T., {Utomo}, D., {et~al.} 2017, \apj, 846, 159,
  \dodoi{10.3847/1538-4357/aa86aa}

\bibitem[{{Broeils} \& {Courteau}(1997)}]{1997ASPC..117...74B}
{Broeils}, A.~H., \& {Courteau}, S. 1997, in Astronomical Society of the
  Pacific Conference Series, Vol. 117, Dark and Visible Matter in Galaxies and
  Cosmological Implications, ed. M.~{Persic} \& P.~{Salucci}, 74.
\newblock \doarXiv{astro-ph/9610264}

\bibitem[{{Bundy} {et~al.}(2015){Bundy}, {Bershady}, {Law}, {Yan}, {Drory},
  {MacDonald}, {Wake}, {Cherinka}, {S{\'a}nchez-Gallego}, {Weijmans}, {Thomas},
  {Tremonti}, {Masters}, {Coccato}, {Diamond-Stanic}, {Arag{\'o}n-Salamanca},
  {Avila-Reese}, {Badenes}, {Falc{\'o}n-Barroso}, {Belfiore}, {Bizyaev},
  {Blanc}, {Bland-Hawthorn}, {Blanton}, {Brownstein}, {Byler}, {Cappellari},
  {Conroy}, {Dutton}, {Emsellem}, {Etherington}, {Frinchaboy}, {Fu}, {Gunn},
  {Harding}, {Johnston}, {Kauffmann}, {Kinemuchi}, {Klaene}, {Knapen},
  {Leauthaud}, {Li}, {Lin}, {Maiolino}, {Malanushenko}, {Malanushenko}, {Mao},
  {Maraston}, {McDermid}, {Merrifield}, {Nichol}, {Oravetz}, {Pan}, {Parejko},
  {Sanchez}, {Schlegel}, {Simmons}, {Steele}, {Steinmetz}, {Thanjavur},
  {Thompson}, {Tinker}, {van den Bosch}, {Westfall}, {Wilkinson}, {Wright},
  {Xiao}, \& {Zhang}}]{2015ApJ...798....7B}
{Bundy}, K., {Bershady}, M.~A., {Law}, D.~R., {et~al.} 2015, \apj, 798, 7,
  \dodoi{10.1088/0004-637X/798/1/7}

\bibitem[{{Bureau} \& {Carignan}(2002)}]{2002AJ....123.1316B}
{Bureau}, M., \& {Carignan}, C. 2002, \aj, 123, 1316, \dodoi{10.1086/338899}

\bibitem[{{Cardelli} {et~al.}(1989){Cardelli}, {Clayton}, \&
  {Mathis}}]{1989ApJ...345..245C}
{Cardelli}, J.~A., {Clayton}, G.~C., \& {Mathis}, J.~S. 1989, \apj, 345, 245,
  \dodoi{10.1086/167900}

\bibitem[{{Dalcanton} \& {Stilp}(2010)}]{2010ApJ...721..547D}
{Dalcanton}, J.~J., \& {Stilp}, A.~M. 2010, \apj, 721, 547,
  \dodoi{10.1088/0004-637X/721/1/547}

\bibitem[{{Davis} {et~al.}(2013){Davis}, {Alatalo}, {Bureau}, {Cappellari},
  {Scott}, {Young}, {Blitz}, {Crocker}, {Bayet}, {Bois}, {Bournaud}, {Davies},
  {de Zeeuw}, {Duc}, {Emsellem}, {Khochfar}, {Krajnovi{\'c}}, {Kuntschner},
  {Lablanche}, {McDermid}, {Morganti}, {Naab}, {Oosterloo}, {Sarzi}, {Serra},
  \& {Weijmans}}]{2013MNRAS.429..534D}
{Davis}, T.~A., {Alatalo}, K., {Bureau}, M., {et~al.} 2013, \mnras, 429, 534,
  \dodoi{10.1093/mnras/sts353}

\bibitem[{{de Blok} {et~al.}(2001){de Blok}, {McGaugh}, \&
  {Rubin}}]{2001AJ....122.2396D}
{de Blok}, W.~J.~G., {McGaugh}, S.~S., \& {Rubin}, V.~C. 2001, \aj, 122, 2396,
  \dodoi{10.1086/323450}

\bibitem[{{de Blok} {et~al.}(2008){de Blok}, {Walter}, {Brinks},
  {Trachternach}, {Oh}, \& {Kennicutt}}]{2008AJ....136.2648D}
{de Blok}, W.~J.~G., {Walter}, F., {Brinks}, E., {et~al.} 2008, \aj, 136, 2648,
  \dodoi{10.1088/0004-6256/136/6/2648}

\bibitem[{{Di Teodoro} \& {Fraternali}(2015)}]{2015MNRAS.451.3021D}
{Di Teodoro}, E.~M., \& {Fraternali}, F. 2015, \mnras, 451, 3021,
  \dodoi{10.1093/mnras/stv1213}

\bibitem[{{Douglass} {et~al.}(2019){Douglass}, {Smith}, \&
  {Demina}}]{2019ApJ...886..153D}
{Douglass}, K.~A., {Smith}, J.~A., \& {Demina}, R. 2019, \apj, 886, 153,
  \dodoi{10.3847/1538-4357/ab4bce}

\bibitem[{{Epinat} {et~al.}(2009){Epinat}, {Contini}, {Le F{\`e}vre},
  {Vergani}, {Garilli}, {Amram}, {Queyrel}, {Tasca}, \&
  {Tresse}}]{2009A&A...504..789E}
{Epinat}, B., {Contini}, T., {Le F{\`e}vre}, O., {et~al.} 2009, \aap, 504, 789,
  \dodoi{10.1051/0004-6361/200911995}

\bibitem[{{Espinosa-Ponce} {et~al.}(2020){Espinosa-Ponce}, {S{\'a}nchez},
  {Morisset}, {Barrera-Ballesteros}, {Galbany}, {Garc{\'\i}a-Benito},
  {Lacerda}, \& {Mast}}]{2020MNRAS.494.1622E}
{Espinosa-Ponce}, C., {S{\'a}nchez}, S.~F., {Morisset}, C., {et~al.} 2020,
  \mnras, 494, 1622, \dodoi{10.1093/mnras/staa782}

\bibitem[{{Haffner} {et~al.}(2009){Haffner}, {Dettmar}, {Beckman}, {Wood},
  {Slavin}, {Giammanco}, {Madsen}, {Zurita}, \&
  {Reynolds}}]{2009RvMP...81..969H}
{Haffner}, L.~M., {Dettmar}, R.~J., {Beckman}, J.~E., {et~al.} 2009, Reviews of
  Modern Physics, 81, 969, \dodoi{10.1103/RevModPhys.81.969}

\bibitem[{{Iorio} {et~al.}(2017){Iorio}, {Fraternali}, {Nipoti}, {Di Teodoro},
  {Read}, \& {Battaglia}}]{2017MNRAS.466.4159I}
{Iorio}, G., {Fraternali}, F., {Nipoti}, C., {et~al.} 2017, \mnras, 466, 4159,
  \dodoi{10.1093/mnras/stw3285}

\bibitem[{{Kelz} {et~al.}(2006){Kelz}, {Verheijen}, {Roth}, {Bauer}, {Becker},
  {Paschke}, {Popow}, {S{\'a}nchez}, \& {Laux}}]{2006PASP..118..129K}
{Kelz}, A., {Verheijen}, M. A.~W., {Roth}, M.~M., {et~al.} 2006, \pasp, 118,
  129, \dodoi{10.1086/497455}

\bibitem[{{Kewley} {et~al.}(2001){Kewley}, {Dopita}, {Sutherland}, {Heisler},
  \& {Trevena}}]{2001ApJ...556..121K}
{Kewley}, L.~J., {Dopita}, M.~A., {Sutherland}, R.~S., {Heisler}, C.~A., \&
  {Trevena}, J. 2001, \apj, 556, 121, \dodoi{10.1086/321545}

\bibitem[{{Kewley} {et~al.}(2006){Kewley}, {Groves}, {Kauffmann}, \&
  {Heckman}}]{2006MNRAS.372..961K}
{Kewley}, L.~J., {Groves}, B., {Kauffmann}, G., \& {Heckman}, T. 2006, \mnras,
  372, 961, \dodoi{10.1111/j.1365-2966.2006.10859.x}

\bibitem[{{Kuzio de Naray} {et~al.}(2008){Kuzio de Naray}, {McGaugh}, \& {de
  Blok}}]{2008ApJ...676..920K}
{Kuzio de Naray}, R., {McGaugh}, S.~S., \& {de Blok}, W.~J.~G. 2008, \apj, 676,
  920, \dodoi{10.1086/527543}

\bibitem[{{Lacerda} {et~al.}(2018){Lacerda}, {Cid Fernandes}, {Couto},
  {Stasi{\'n}ska}, {Garc{\'\i}a-Benito}, {Vale Asari}, {P{\'e}rez},
  {Gonz{\'a}lez Delgado}, {S{\'a}nchez}, \& {de Amorim}}]{2018MNRAS.474.3727L}
{Lacerda}, E.~A.~D., {Cid Fernandes}, R., {Couto}, G.~S., {et~al.} 2018,
  \mnras, 474, 3727, \dodoi{10.1093/mnras/stx3022}

\bibitem[{{Lang} {et~al.}(2020){Lang}, {Meidt}, {Rosolowsky}, {Nofech},
  {Schinnerer}, {Leroy}, {Emsellem}, {Pessa}, {Glover}, {Groves}, {Hughes},
  {Kruijssen}, {Querejeta}, {Schruba}, {Bigiel}, {Blanc}, {Chevance},
  {Colombo}, {Faesi}, {Henshaw}, {Herrera}, {Liu}, {Pety}, {Puschnig}, {Saito},
  {Sun}, \& {Usero}}]{2020ApJ...897..122L}
{Lang}, P., {Meidt}, S.~E., {Rosolowsky}, E., {et~al.} 2020, \apj, 897, 122,
  \dodoi{10.3847/1538-4357/ab9953}

\bibitem[{{Law} {et~al.}(2016){Law}, {Cherinka}, {Yan}, {Andrews}, {Bershady},
  {Bizyaev}, {Blanc}, {Blanton}, {Bolton}, {Brownstein}, {Bundy}, {Chen},
  {Drory}, {D'Souza}, {Fu}, {Jones}, {Kauffmann}, {MacDonald}, {Masters},
  {Newman}, {Parejko}, {S{\'a}nchez-Gallego}, {S{\'a}nchez}, {Schlegel},
  {Thomas}, {Wake}, {Weijmans}, {Westfall}, \& {Zhang}}]{2016AJ....152...83L}
{Law}, D.~R., {Cherinka}, B., {Yan}, R., {et~al.} 2016, \aj, 152, 83,
  \dodoi{10.3847/0004-6256/152/4/83}

\bibitem[{{Law} {et~al.}(2021){Law}, {Westfall}, {Bershady}, {Cappellari},
  {Yan}, {Belfiore}, {Bizyaev}, {Brownstein}, {Chen}, {Cherinka}, {Drory},
  {Lazarz}, \& {Shetty}}]{2021AJ....161...52L}
{Law}, D.~R., {Westfall}, K.~B., {Bershady}, M.~A., {et~al.} 2021, \aj, 161,
  52, \dodoi{10.3847/1538-3881/abcaa2}

\bibitem[{{Leung} {et~al.}(2018){Leung}, {Leaman}, {van de Ven}, {Lyubenova},
  {Zhu}, {Bolatto}, {Falc{\'o}n-Barroso}, {Blitz}, {Dannerbauer}, {Fisher},
  {Levy}, {Sanchez}, {Utomo}, {Vogel}, {Wong}, \&
  {Ziegler}}]{2018MNRAS.477..254L}
{Leung}, G. Y.~C., {Leaman}, R., {van de Ven}, G., {et~al.} 2018, \mnras, 477,
  254, \dodoi{10.1093/mnras/sty288}

\bibitem[{{Levy} {et~al.}(2018){Levy}, {Bolatto}, {Teuben}, {S{\'a}nchez},
  {Barrera-Ballesteros}, {Blitz}, {Colombo}, {Garc{\'\i}a-Benito},
  {Herrera-Camus}, {Husemann}, {Kalinova}, {Lan}, {Leung}, {Mast}, {Utomo},
  {van de Ven}, {Vogel}, \& {Wong}}]{2018ApJ...860...92L}
{Levy}, R.~C., {Bolatto}, A.~D., {Teuben}, P., {et~al.} 2018, \apj, 860, 92,
  \dodoi{10.3847/1538-4357/aac2e5}

\bibitem[{{Levy} {et~al.}(2019){Levy}, {Bolatto}, {S{\'a}nchez}, {Blitz},
  {Colombo}, {Kalinova}, {L{\'o}pez-Cob{\'a}}, {Ostriker}, {Teuben}, {Utomo},
  {Vogel}, \& {Wong}}]{2019ApJ...882...84L}
{Levy}, R.~C., {Bolatto}, A.~D., {S{\'a}nchez}, S.~F., {et~al.} 2019, \apj,
  882, 84, \dodoi{10.3847/1538-4357/ab2ed4}

\bibitem[{{Lin} {et~al.}(2020){Lin}, {Ellison}, {Pan}, {Thorp}, {Su},
  {S{\'a}nchez}, {Belfiore}, {Bothwell}, {Bundy}, {Chen}, {Concas}, {Hsieh},
  {Hsieh}, {Li}, {Maiolino}, {Masters}, {Newman}, {Rowlands}, {Shi},
  {Smethurst}, {Stark}, {Xiao}, \& {Yu}}]{2020ApJ...903..145L}
{Lin}, L., {Ellison}, S.~L., {Pan}, H.-A., {et~al.} 2020, \apj, 903, 145,
  \dodoi{10.3847/1538-4357/abba3a}

\bibitem[{{{\"O}stlin} {et~al.}(1999){{\"O}stlin}, {Amram}, {Masegosa},
  {Bergvall}, \& {Boulesteix}}]{1999A&AS..137..419O}
{{\"O}stlin}, G., {Amram}, P., {Masegosa}, J., {Bergvall}, N., \& {Boulesteix},
  J. 1999, \aaps, 137, 419, \dodoi{10.1051/aas:1999488}

\bibitem[{{Rogstad} {et~al.}(1974){Rogstad}, {Lockhart}, \&
  {Wright}}]{1974ApJ...193..309R}
{Rogstad}, D.~H., {Lockhart}, I.~A., \& {Wright}, M.~C.~H. 1974, \apj, 193,
  309, \dodoi{10.1086/153164}

\bibitem[{{Rosa-Gonz{\'a}lez} {et~al.}(2002){Rosa-Gonz{\'a}lez}, {Terlevich},
  \& {Terlevich}}]{2002MNRAS.332..283R}
{Rosa-Gonz{\'a}lez}, D., {Terlevich}, E., \& {Terlevich}, R. 2002, \mnras, 332,
  283, \dodoi{10.1046/j.1365-8711.2002.05285.x}

\bibitem[{{S{\'a}nchez}(2020)}]{2020ARA&A..58...99S}
{S{\'a}nchez}, S.~F. 2020, \araa, 58, 99,
  \dodoi{10.1146/annurev-astro-012120-013326}

\bibitem[{{S{\'a}nchez} {et~al.}(2012){S{\'a}nchez}, {Kennicutt}, {Gil de Paz},
  {van de Ven}, {V{\'\i}lchez}, {Wisotzki}, {Walcher}, {Mast}, {Aguerri},
  {Albiol-P{\'e}rez}, {Alonso-Herrero}, {Alves}, {Bakos}, {Bart{\'a}kov{\'a}},
  {Bland-Hawthorn}, {Boselli}, {Bomans}, {Castillo-Morales}, {Cortijo-Ferrero},
  {de Lorenzo-C{\'a}ceres}, {Del Olmo}, {Dettmar}, {D{\'\i}az}, {Ellis},
  {Falc{\'o}n-Barroso}, {Flores}, {Gallazzi}, {Garc{\'\i}a-Lorenzo},
  {Gonz{\'a}lez Delgado}, {Gruel}, {Haines}, {Hao}, {Husemann},
  {Igl{\'e}sias-P{\'a}ramo}, {Jahnke}, {Johnson}, {Jungwiert}, {Kalinova},
  {Kehrig}, {Kupko}, {L{\'o}pez-S{\'a}nchez}, {Lyubenova}, {Marino},
  {M{\'a}rmol-Queralt{\'o}}, {M{\'a}rquez}, {Masegosa}, {Meidt},
  {Mendez-Abreu}, {Monreal-Ibero}, {Montijo}, {Mour{\~a}o}, {Palacios-Navarro},
  {Papaderos}, {Pasquali}, {Peletier}, {P{\'e}rez}, {P{\'e}rez}, {Quirrenbach},
  {Rela{\~n}o}, {Rosales-Ortega}, {Roth}, {Ruiz-Lara},
  {S{\'a}nchez-Bl{\'a}zquez}, {Sengupta}, {Singh}, {Stanishev}, {Trager},
  {Vazdekis}, {Viironen}, {Wild}, {Zibetti}, \&
  {Ziegler}}]{2012A&A...538A...8S}
{S{\'a}nchez}, S.~F., {Kennicutt}, R.~C., {Gil de Paz}, A., {et~al.} 2012,
  \aap, 538, A8, \dodoi{10.1051/0004-6361/201117353}

\bibitem[{{S{\'a}nchez} {et~al.}(2016{\natexlab{a}}){S{\'a}nchez}, {P{\'e}rez},
  {S{\'a}nchez-Bl{\'a}zquez}, {Gonz{\'a}lez}, {Ros{\'a}les-Ortega},
  {Cano-D{\'\i}az}, {L{\'o}pez-Cob{\'a}}, {Marino}, {Gil de Paz}, {Moll{\'a}},
  {L{\'o}pez-S{\'a}nchez}, {Ascasibar}, \&
  {Barrera-Ballesteros}}]{2016RMxAA..52...21S}
{S{\'a}nchez}, S.~F., {P{\'e}rez}, E., {S{\'a}nchez-Bl{\'a}zquez}, P., {et~al.}
  2016{\natexlab{a}}, \rmxaa, 52, 21.
\newblock \doarXiv{1509.08552}

\bibitem[{{S{\'a}nchez} {et~al.}(2016{\natexlab{b}}){S{\'a}nchez}, {P{\'e}rez},
  {S{\'a}nchez-Bl{\'a}zquez}, {Garc{\'\i}a-Benito}, {Ibarra-Mede},
  {Gonz{\'a}lez}, {Rosales-Ortega}, {S{\'a}nchez-Menguiano}, {Ascasibar},
  {Bitsakis}, {Law}, {Cano-D{\'\i}az}, {L{\'o}pez-Cob{\'a}}, {Marino}, {Gil de
  Paz}, {L{\'o}pez-S{\'a}nchez}, {Barrera-Ballesteros}, {Galbany}, {Mast},
  {Abril-Melgarejo}, \& {Roman-Lopes}}]{2016RMxAA..52..171S}
---. 2016{\natexlab{b}}, \rmxaa, 52, 171.
\newblock \doarXiv{1602.01830}

\bibitem[{{S{\'a}nchez} {et~al.}(2016{\natexlab{c}}){S{\'a}nchez},
  {Garc{\'\i}a-Benito}, {Zibetti}, {Walcher}, {Husemann}, {Mendoza}, {Galbany},
  {Falc{\'o}n-Barroso}, {Mast}, {Aceituno}, {Aguerri}, {Alves}, {Amorim},
  {Ascasibar}, {Barrado-Navascues}, {Barrera-Ballesteros}, {Bekerait{\`e}},
  {Bland-Hawthorn}, {Cano D{\'\i}az}, {Cid Fernandes}, {Cavichia}, {Cortijo},
  {Dannerbauer}, {Demleitner}, {D{\'\i}az}, {Dettmar}, {de
  Lorenzo-C{\'a}ceres}, {del Olmo}, {Galazzi}, {Garc{\'\i}a-Lorenzo}, {Gil de
  Paz}, {Gonz{\'a}lez Delgado}, {Holmes}, {Igl{\'e}sias-P{\'a}ramo}, {Kehrig},
  {Kelz}, {Kennicutt}, {Kleemann}, {Lacerda}, {L{\'o}pez Fern{\'a}ndez},
  {L{\'o}pez S{\'a}nchez}, {Lyubenova}, {Marino}, {M{\'a}rquez},
  {Mendez-Abreu}, {Moll{\'a}}, {Monreal-Ibero}, {Ortega Minakata},
  {Torres-Papaqui}, {P{\'e}rez}, {Rosales-Ortega}, {Roth},
  {S{\'a}nchez-Bl{\'a}zquez}, {Schilling}, {Spekkens}, {Vale Asari}, {van den
  Bosch}, {van de Ven}, {Vilchez}, {Wild}, {Wisotzki}, {Y{\i}ld{\i}r{\i}m}, \&
  {Ziegler}}]{2016A&A...594A..36S}
{S{\'a}nchez}, S.~F., {Garc{\'\i}a-Benito}, R., {Zibetti}, S., {et~al.}
  2016{\natexlab{c}}, \aap, 594, A36, \dodoi{10.1051/0004-6361/201628661}

\bibitem[{{S{\'a}nchez} {et~al.}(2018){S{\'a}nchez}, {Avila-Reese},
  {Hernandez-Toledo}, {Cortes-Su{\'a}rez}, {Rodr{\'\i}guez-Puebla},
  {Ibarra-Medel}, {Cano-D{\'\i}az}, {Barrera-Ballesteros}, {Negrete},
  {Calette}, {de Lorenzo-C{\'a}ceres}, {Ortega-Minakata}, {Aquino},
  {Valenzuela}, {Clemente}, {Storchi-Bergmann}, {Riffel}, {Schimoia}, {Riffel},
  {Rembold}, {Brownstein}, {Pan}, {Yates}, {Mallmann}, \&
  {Bitsakis}}]{2018RMxAA..54..217S}
{S{\'a}nchez}, S.~F., {Avila-Reese}, V., {Hernandez-Toledo}, H., {et~al.} 2018,
  \rmxaa, 54, 217.
\newblock \doarXiv{1709.05438}

\bibitem[{{Simon} {et~al.}(2005){Simon}, {Bolatto}, {Leroy}, {Blitz}, \&
  {Gates}}]{2005ApJ...621..757S}
{Simon}, J.~D., {Bolatto}, A.~D., {Leroy}, A., {Blitz}, L., \& {Gates}, E.~L.
  2005, \apj, 621, 757, \dodoi{10.1086/427684}

\bibitem[{{Sofue} \& {Rubin}(2001)}]{2001ARA&A..39..137S}
{Sofue}, Y., \& {Rubin}, V. 2001, \araa, 39, 137,
  \dodoi{10.1146/annurev.astro.39.1.137}

\bibitem[{{Swaters} {et~al.}(2009){Swaters}, {Sancisi}, {van Albada}, \& {van
  der Hulst}}]{2009A&A...493..871S}
{Swaters}, R.~A., {Sancisi}, R., {van Albada}, T.~S., \& {van der Hulst}, J.~M.
  2009, \aap, 493, 871, \dodoi{10.1051/0004-6361:200810516}

\bibitem[{{van Albada} {et~al.}(1985){van Albada}, {Bahcall}, {Begeman}, \&
  {Sancisi}}]{1985ApJ...295..305V}
{van Albada}, T.~S., {Bahcall}, J.~N., {Begeman}, K., \& {Sancisi}, R. 1985,
  \apj, 295, 305, \dodoi{10.1086/163375}

\bibitem[{{van den Bosch} {et~al.}(2000){van den Bosch}, {Robertson},
  {Dalcanton}, \& {de Blok}}]{2000AJ....119.1579V}
{van den Bosch}, F.~C., {Robertson}, B.~E., {Dalcanton}, J.~J., \& {de Blok},
  W.~J.~G. 2000, \aj, 119, 1579, \dodoi{10.1086/301315}

\bibitem[{{Walcher} {et~al.}(2014){Walcher}, {Wisotzki}, {Bekerait{\'e}},
  {Husemann}, {Iglesias-P{\'a}ramo}, {Backsmann}, {Barrera Ballesteros},
  {Catal{\'a}n-Torrecilla}, {Cortijo}, {del Olmo}, {Garcia Lorenzo},
  {Falc{\'o}n-Barroso}, {Jilkova}, {Kalinova}, {Mast}, {Marino},
  {M{\'e}ndez-Abreu}, {Pasquali}, {S{\'a}nchez}, {Trager}, {Zibetti},
  {Aguerri}, {Alves}, {Bland-Hawthorn}, {Boselli}, {Castillo Morales}, {Cid
  Fernandes}, {Flores}, {Galbany}, {Gallazzi}, {Garc{\'\i}a-Benito}, {Gil de
  Paz}, {Gonz{\'a}lez-Delgado}, {Jahnke}, {Jungwiert}, {Kehrig}, {Lyubenova},
  {M{\'a}rquez Perez}, {Masegosa}, {Monreal Ibero}, {P{\'e}rez}, {Quirrenbach},
  {Rosales-Ortega}, {Roth}, {Sanchez-Blazquez}, {Spekkens}, {Tundo}, {van de
  Ven}, {Verheijen}, {Vilchez}, \& {Ziegler}}]{2014A&A...569A...1W}
{Walcher}, C.~J., {Wisotzki}, L., {Bekerait{\'e}}, S., {et~al.} 2014, \aap,
  569, A1, \dodoi{10.1051/0004-6361/201424198}

\bibitem[{{Westfall} {et~al.}(2019){Westfall}, {Cappellari}, {Bershady},
  {Bundy}, {Belfiore}, {Ji}, {Law}, {Schaefer}, {Shetty}, {Tremonti}, {Yan},
  {Andrews}, {Brownstein}, {Cherinka}, {Coccato}, {Drory}, {Maraston},
  {Parikh}, {S{\'a}nchez-Gallego}, {Thomas}, {Weijmans}, {Barrera-Ballesteros},
  {Du}, {Goddard}, {Li}, {Masters}, {Ibarra Medel}, {S{\'a}nchez}, {Yang},
  {Zheng}, \& {Zhou}}]{2019AJ....158..231W}
{Westfall}, K.~B., {Cappellari}, M., {Bershady}, M.~A., {et~al.} 2019, \aj,
  158, 231, \dodoi{10.3847/1538-3881/ab44a2}

\bibitem[{{Zhu} {et~al.}(2018){Zhu}, {van de Ven}, {van den Bosch}, {Rix},
  {Lyubenova}, {Falc{\'o}n-Barroso}, {Martig}, {Mao}, {Xu}, {Jin}, {Obreja},
  {Grand}, {Dutton}, {Macci{\`o}}, {G{\'o}mez}, {Walcher},
  {Garc{\'\i}a-Benito}, {Zibetti}, \& {S{\'a}nchez}}]{2018NatAs...2..233Z}
{Zhu}, L., {van de Ven}, G., {van den Bosch}, R., {et~al.} 2018, Nature
  Astronomy, 2, 233, \dodoi{10.1038/s41550-017-0348-1}

\end{thebibliography}
\bibliographystyle{aasjournal}



\end{document}